\documentclass[12pt,preprint]{aastex}

\usepackage[bf,small]{caption}

\newcommand{\microns}{$\mu$m}
\def\EW{$W_{\lambda}$}
\def\deg{$^{\circ}$}
\def\ga{\mathrel{\hbox{\rlap{\hbox{\lower4pt\hbox{$\sim$}}}\hbox{$>$}}}}
\def\la{\mathrel{\hbox{\rlap{\hbox{\lower4pt\hbox{$\sim$}}}\hbox{$<$}}}}
\def\msunyr{$M$ \mbox{$_{\normalsize\odot}$}\rm{yr}$^{-1}$}
\def\msun{$M$\mbox{$_{\normalsize\odot}$}}

\def\lsun{$L$\mbox{$_{\normalsize\odot}$}}
\def\kms{\,km~s$^{-1}$}
\def\EW{$W_{\lambda}$}
\def\arcsec{$^{\prime \prime}$}
\def\arcmin{$^{\prime}$}

\slugcomment{}

\shorttitle{A massive RSG cluster in the Scutum-Crux arm}
\shortauthors{Davies et al.}

\begin{document}


\title{A massive cluster of Red Supergiants at the base
  of the Scutum-Crux arm}


\author{Ben Davies\altaffilmark{1}, Don F. Figer\altaffilmark{1},
Rolf-Peter Kudritzki\altaffilmark{2}, John
MacKenty\altaffilmark{3}, Francisco Najarro\altaffilmark{4} and
Artemio Herrero\altaffilmark{5} }

\affil{$^{1}$Chester F.\ Carlson Center for Imaging Science, Rochester
Institute of Technology, 54 Lomb Memorial Drive, Rochester NY, 14623,
USA} 
\affil{$^{2}$Institute for Astronomy, University of Hawaii, 2680
Woodlawn Drive, Honolulu, HI, 96822, USA} 
\affil{$^{3}$Space
Telescope Science Institute, 3700 San Martin Drive, Baltimore, MD,
21218, USA}
\affil{$^{4}$Instituto de Estructura de la Materia, Consejo Superior
  de Investigaciones Cientificas, Calle Serrano 121, 28006 Madrid,
  Spain.} 
\affil{$^{5}$Instituto de Astrofísica de Canarias, Via L\`{a}ctea S/N,
  E-38200 La Laguna, Tenerife, Spain} 







\begin{abstract}
We report on the unprecedented Red Supergiant (RSG) population of a
massive young cluster, located at the base of the Scutum-Crux Galactic
arm. We identify candidate cluster RSGs based on {\it 2MASS}
photometry and medium resolution spectroscopy. With follow-up
high-resolution spectroscopy, we use CO-bandhead equivalent width and
high-precision radial velocity measurements to identify a core
grouping of 26 physically-associated RSGs -- the largest such cluster
known to-date. Using the stars' velocity dispersion, and their
inferred luminosities in conjuction with evolutionary models, we argue
that the cluster has an initial mass of $\sim$40,000\msun, and is
therefore among the most massive in the galaxy. Further, the cluster
is only a few hundred parsecs away from the cluster of 14 RSGs
recently reported by Figer et al (2006). These two RSG clusters
represent 20\% of all known RSGs in the Galaxy, and now offer the
unique opportunity to study the pre-supernova evolution of massive
stars, and the Blue- to Red-Supergiant ratio at uniform
metallicity. We use GLIMPSE, MIPSGAL and MAGPIS survey data to
identify several objects in the field of the larger cluster which seem
to be indicative of recent region-wide starburst activity at the point
where the Scutum-Crux arm intercepts the Galactic bulge. Future
abundance studies of these clusters will therefore permit the study of
the chemical evolution and metallicity gradient of the Galaxy in the
region where the disk meets the bulge.
\end{abstract}



\section{Introduction}
Massive stars play a pivotal role in the evolution of their host
galaxies. As main-sequence O stars, they emit copious amounts of
ionizing UV radiation. Their post-MS evolution is characterized by
brief but extreme mass-losing episodes, such as the Red Supergiant
(RSG), Luminous Blue Variable (LBV) and Wolf-Rayet (WR) phases, during
which they inject the interstellar medium (ISM) with mechanical energy
and chemically-processed material. When they end their lives as
core-collapse supernovae (SNe), they inject the ISM with heavy
elements and drive shocks into their surroundings, strongly
influencing subsequent local star-formation.

The stellar end state, i.e.\ neutron star, black-hole, or complete
disruption, depends on the terminal mass of the star
\citep{Heger03}. Additionally, the appearance of the SN explosion is
thought to be linked to the progenitor. The hydrogen-poor Type Ib+c SN
are thought to have WR progenitors, while the progenitors of many of
the H-rich Type II-P have been identified as RSGs from archival {\it
HST} images \citep{vD03, Smartt04, Maund04}. However, the most
clear-cut case of a Type II-P progenitor remains SN1987A, which was a
Blue Supergiant (BSG) \citep{Sonneborn87}.

Predicting the evolution of massive stars from post-main sequence to
the end of their lives is notoriously problematic. It is driven by the
star's mass-loss behaviour, which in turn is strongly dependent on
factors such as metallicity and rotation which are poorly constrained
\citep[see review of][]{K-P00}. Meanwhile, empirical studies of
massive stellar evolution are hampered by low-number statistics, due
to the steepness of the Initial Mass Function (IMF), the short
lifetimes of the stars, and the obcurring effect of the gas and dust
in the Galactic plane.

Galactic young massive clusters provide us with the ideal natural
laboratories in which to study massive stellar evolution. Such objects
provide a coeval sample of massive stars under the constraint of
uniform metallicity, whilst being close enough to resolve the
individual stars. Unfortunately, such objects are rare. Until
recently, only Westerlund~1 \citep[Wd~1, ][]{Clark05}, the Arches
Cluster \citep{Figer02}, the Quintuplet Cluster \citep{Figer99}, and
the Galactic Centre (GC) cluster \citep{Figer04} were known to be
massive enough and young enough to harbour statistically-significant
numbers of massive stars. Ages of these clusters range from $\sim$3Myr
(Arches) to $\sim$5Myr (Wd 1). Hence, while they are both young and
massive enough to contain large numbers of O stars and WRs, they are
{\it too} young to have similar numbers of RSGs, which are expected
after $\sim$6Myr.

Using catalogues of Galactic Plane cluster candidates
\citep{Bica03a,Bica03b,Dutra03}, \citet[][ hereafter FMR06]{Figer06}
made the discovery of an unprecedented cluster of 14 RSGs at a
Galactic longitude of $l=25$\degr, hereafter known as RSGC1. At the
time of discovery this object contained by far the greatest number of
RSGs of all known Galactic clusters, a record previously held by
NGC~7419 with 5 \citep{Caron03}.

RSGC1 is located at the base of the Scutum-Crux arm, close to where it
meets the Galactic bulge. It is separated by $\sim$1\deg\ from another
reddened cluster, Stephenson 2. In the discovery paper,
\citet{Stephenson90} speculated that the cluster may harbour several
RSGs, possibly up to 10, based on the brightness of the stars in the
$I$ band. The cluster was also studied by \citet{Nakaya01} and
\citet{Ortolani02}, who estimated distances of 1.5kpc/5.9kpc, and ages
of 50Myr/20Myr respectively, from optical and infrared photometry.

Here, we present low- and high-resolution spectroscopy of over 40 red
stars in this cluster, and combine this with {\it 2MASS}, {\it MSX}
and {\it GLIMPSE} photometry. We show from high-precision radial
velocity measurements and IR photometry that the cluster, hereafter
known as RSGC2, is host to 26 RSGs, by far the largest known
population in the Galaxy. We use this velocity information, in
conjuction with the stars' spectra and stellar evolution models to
better constrain the distance and age of the cluster.

Between them RSGC1 and 2 are host to $\sim$20\% of all known RSGs in
the Galaxy, and now offer us the opportunity to study a coeval sample
of Type II-P SN progenitors, as well as the BSG/RSG ratio important in
constraining stellar evolutionary models, at uniform
metallicity. Further, the location of the clusters within the Galaxy
will allow future metallicity studies to probe the apparent chemical
discontinuity observed to exist between the Galactic disk and bulge
\citep[see e.g.][]{Ramirez00,Smartt01,Najarro04}. 


We begin in Sect.\ \ref{sec:obs} with a description of our
observations, including target selection, data reduction and analysis
techniques. We present the results of the data analysis in Sect.\
\ref{sec:results}, and argue which of the stars observed are members
of the cluster. In Sect.\ \ref{sec:disc} we estimate the cluster age
and mass, before discussing the two remarkable Scutum-Crux RSGCs in
the context of of other massive Galactic clusters, and their
significance in the study of various aspects of stellar/galactic
evolution.

\section{Observations \& data reduction} \label{sec:obs}
\subsection{Target selection \& strategy}
For a coeval population of stars, RSGs are typically $\sim$4 mags
brighter in the $K$-band than any other stars in the
cluster. Therefore, in order to identify potential RSGs, we compliled
a list of candidate stars within an 7\arcmin\ radius of the cluster
centre \citep[as defined by][]{Stephenson90} based on their
$K_{S}$-band magnitudes in the Two-Micron All Sky Survey ({\it 2MASS})
point-source catalogue \citep{Cutri03}.

A key spectral diagnostic for late-type stars is the CO bandhead
feature at 2.295\microns. As shown in FMR06, the feature is evident in
spectral-type G and later, and is extremely prominant in M-type
stars. Further, the feature is stronger in supergiants than giants and
dwarfs. To identify those stars with CO absorption, we observed the
brightest 50 stars in our target list, as well as others, with {\it
IRMOS} -- the Infra-Red Multi-Object Spectrograph \citep{MacKenty03} --
at the KPNO-4m during April 2006. 

It is likely that this CO sub-sample of stars is contaminated by
foreground/unrelated M dwarfs and giants. To determine which of our
sample are physically related, we obtained follow-up high-resolution
spectroscopic data of the CO feature to measure high-precision radial
velocities of the stars. These data were taken with {\it NIRSPEC}, the
cross-dispersed echelle spectrograph at the Keck II telescope
\citep{McLean98}, during two observing runs in May and August 2006. In
total we observed all but one of the brightest 33 stars, as well as 11
others. As will be shown in Sect.\ \ref{sec:results}, the fainter
stars of the sample are more likely to be foreground objects and not
supergiants, therefore the few stars we missed will not have a
significant impact on our conclusions.

\begin{figure}[p]
\centering
\includegraphics[width=12cm,bb=60 20 700 600]{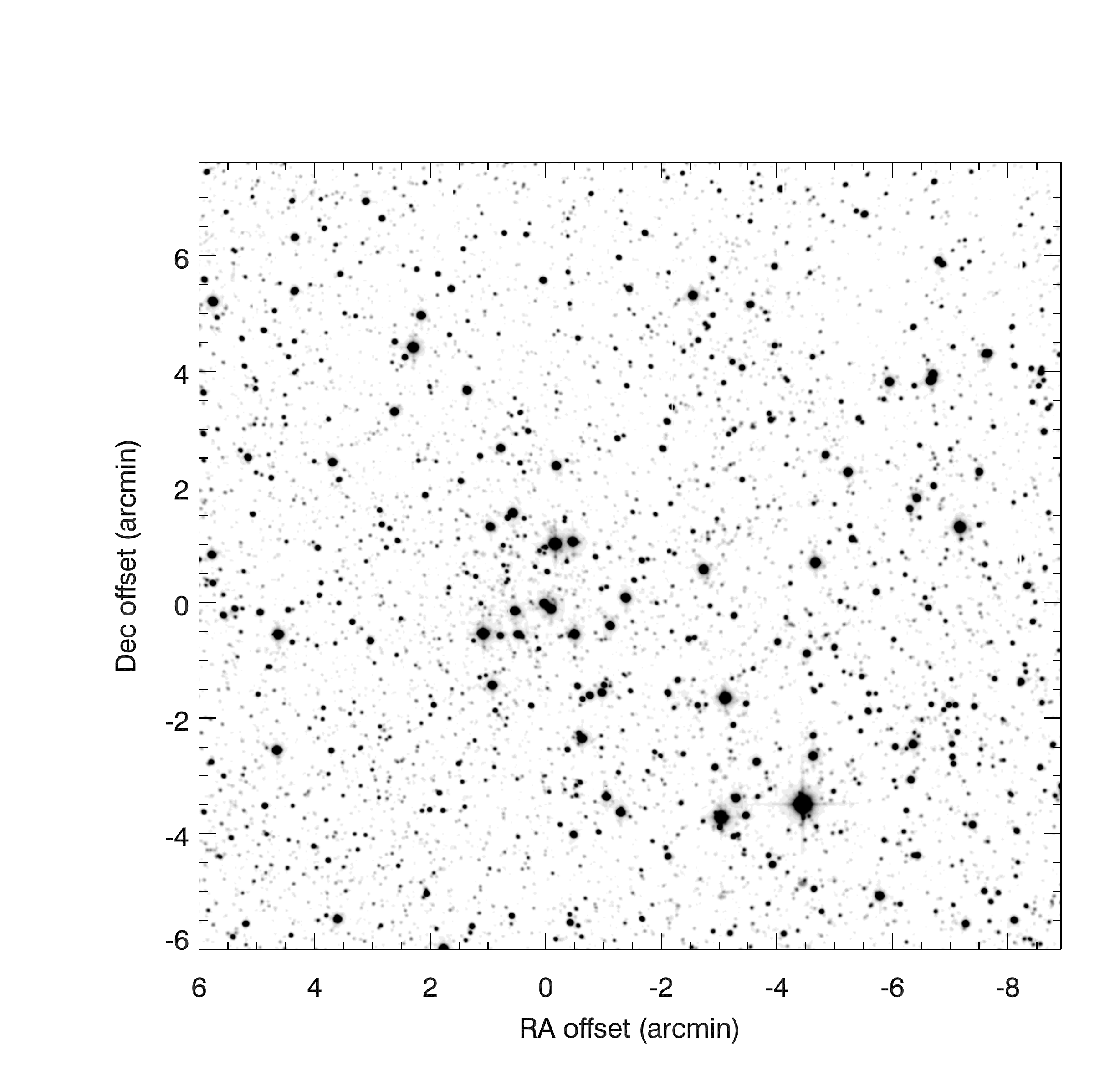}
\includegraphics[width=12cm,bb=40 30 700 640]{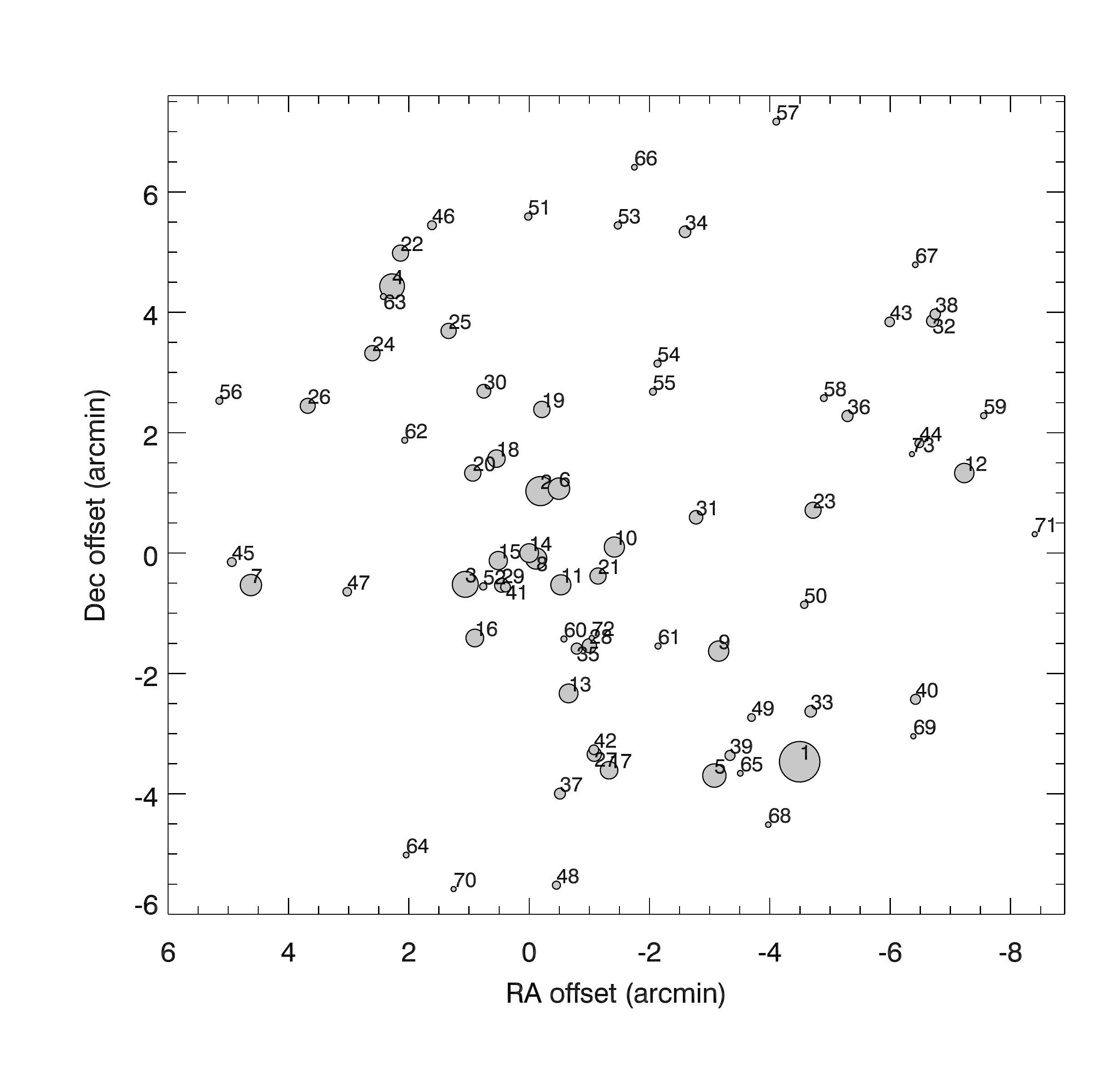}
  \caption{2MASS K-band image of RSGC2 ({\it upper}), and finding
  chart ({\it lower}). Coordinates are centered on 18 39 20.4, -6 01
  41 (2000), following \citet{Stephenson90}. The sizes of the plotting
  symbols scale linearly with K-band magnitude.}
  \label{fig:kimage}
\end{figure}

Table \ref{tab:obsdata} lists the brightest 72 stars within a
7\arcmin\ radius of the cluster centre, together with their {\it
2MASS} photometry and the dates observed. The stars are indexed in
order of their $K_{S}$-band magnitude; also listed are any
identification in \citet{Stephenson90} and \citet{Nakaya01}. A {\it
2MASS} $K_{S}$-band image of RSGC2 is shown in Fig.~\ref{fig:kimage}a,
and Fig.~\ref{fig:kimage}b shows a finding chart for the stars. Both the
K-band image and the finding chart are centered on the approximate
center of the cluster, $18^{h} 39' 20.4''$, -6\deg $01' 41''$, epoch
2000 (Star 5 in S90, Star 14 in this paper).

\begin{table}[p]
    \caption{Observational data for the brightest stars in the {\it
    2MASS} point-source catalogue within 7\arcmin\ of the cluster
    center. Star are indexed according to their $K_{S}$-band
    magnitudes (column 1). Columns 2 and 3 denote identifications in
    the previous studies of \citet{Stephenson90} and \citet{Nakaya01}.
    Columns 4 and 5 list the J2000 coordinates of each star, and the
    {\it 2MASS} photometry is listed in columns 6-8. The {\it NIRSPEC}
    observation date is listed in the final column. Note that Star 32b
    is not resolved in {\it 2MASS} due to its proximity to Star 32. }
    \begin{tabular}{lccccccccc}
      \hline\hline
      ID&S90&N01&RA&Dec&J&H&$K_{s}$& \multicolumn{2}{c}{Obs Date}\\
      & & &\multicolumn{2}{c}{(J2000)} & & & & {\it NIRSPEC} & {\it IRMOS}\\
      \hline
      1&&&18 39 02.4&-06 05 10.6&7.15&4.698&2.900&08/12/06&04/17/06\\
      2&2&&18 39 19.6&-06 00 40.8&6.899&5.045&4.117&05/05/06&\\
      3&10&&18 39 24.6&-06 02 13.8&7.273&5.458&4.499&05/05/06&\\
      4&&&18 39 29.5&-05 57 16.6&7.906&5.705&4.647&08/12/06&04/18/06\\
      5&&&18 39 08.1&-06 05 24.4&8.532&6.054&4.822&08/12/06&04/17/06\\
      6&1&&18 39 18.4&-06 00 38.4&7.717&5.919&5.062&05/05/06&\\
      7&&&18 39 38.9&-06 02 14.5&7.585&5.867&5.090&08/12/06&04/20/06\\
      8&4&&18 39 19.9&-06 01 48.1&7.817&6.015&5.106&05/05/06&\\
      9&&&18 39 06.8&-06 03 20.3&8.569&6.308&5.233&08/12/06&04/17/06\\
      10&&1102&18 39 14.7&-06 01 36.6&8.218&6.214&5.244&05/05/06&\\
      11&&1180&18 39 18.3&-06 02 14.3&8.354&6.207&5.256&05/05/06&\\
      12&&&18 38 51.4&-06 00 22.8&7.221&5.943&5.354&-&\\
      13&&1230&18 39 17.7&-06 04 02.5&8.421&6.387&5.439&08/12/06&04/17/06\\
      14&5&&18 39 20.4&-06 01 42.6&8.222&6.355&5.443&05/05/06&\\
      15&6&&18 39 22.4&-06 01 50.1&8.129&6.346&5.513&05/05/06&\\
      16&8&&18 39 24.0&-06 03 07.3&8.235&6.444&5.597&05/05/06&\\
      17&&&18 39 15.1&-06 05 19.1&8.709&6.613&5.619&08/12/06&04/17/06\\
      18&7&&18 39 22.5&-06 00 08.4&8.179&6.451&5.632&08/12/06&\\
      19&3&&18 39 19.5&-05 59 19.4&8.282&6.584&5.801&08/12/06&\\
      20&9&&18 39 24.1&-06 00 22.8&8.426&6.656&5.805&08/12/06&\\
      21&&598&18 39 15.8&-06 02 05.5&9.115&6.925&5.824&05/05/06&\\
      22&&&18 39 28.9&-05 56 43.6&9.142&7.071&5.825&08/12/06&04/18/06\\
      23&&&18 39 01.5&-06 00 59.9&10.088&7.219&5.840&08/13/06&04/20/06\\
      24&&&18 39 30.8&-05 58 23.3&7.356&6.368&5.960&08/12/06&04/17/06\\
      25&&&18 39 25.7&-05 58 01.1&8.911&6.965&5.975&08/12/06&04/17/06\\
      26&&&18 39 35.1&-05 59 15.8&8.676&6.902&6.003&08/12/06&04/17/06\\
      27&&&18 39 16.0&-06 05 03.2&9.058&7.055&6.129&08/12/06&04/17/06\\
      28&&&18 39 16.4&-06 03 15.0&7.626&6.589&6.132&08/12/06&04/17/06\\
      29&&&18 39 22.2&-06 02 14.7&8.608&6.877&6.146&05/05/06&\\
      30&&&18 39 23.4&-05 59 01.3&8.711&6.956&6.200&08/12/06&\\
      31&&978&18 39 09.3&-06 01 06.9&9.373&7.232&6.244&08/13/06&04/20/06\\
      32&&&18 38 53.5&-05 57 51.2&9.676&7.571&6.490&08/13/06&\\
      32b&&&18 38 52.8&-05 57 40.0&-&-&-&08/13/06&\\
      \hline
    \end{tabular}
    \label{tab:obsdata}
\end{table}
\addtocounter{table}{-1}
\begin{table}[p]
  \caption{Cont.}
    \begin{tabular}{lccccccccc}
      \hline\hline
      ID&S90&N01&RA&Dec&J&H&$K_{s}$& \multicolumn{2}{c}{Obs Date}\\
      & & &\multicolumn{2}{c}{(J2000)} & & & & {\it NIRSPEC} & {\it IRMOS}\\
      \hline
      33&&&18 39 01.6&-06 04 20.4&10.359&7.861&6.581&08/12/06&04/17/06\\
      34&&&18 39 10.0&-05 56 22.4&9.691&7.579&6.585&-&\\
      35&&&18 39 17.2&-06 03 17.9&8.878&7.291&6.651&08/12/06&04/17/06\\
      36&&&18 38 59.2&-05 59 26.0&11.133&8.142&6.655&08/13/06&\\
      37&&&18 39 18.3&-06 05 42.4&10.177&7.843&6.693&-&\\
      38&&&18 38 53.4&-05 57 44.4&8.572&7.324&6.794&08/13/06&\\
      39&&&18 39 07.0&-06 05 04.4&9.695&7.764&6.846&-&04/17/06\\
      40&&&18 38 54.7&-06 04 08.4&10.240&7.929&6.862&-&\\
      41&&&18 39 21.9&-06 02 16.5&9.452&7.781&6.904&-&\\
      42&&&18 39 16.1&-06 04 58.7&12.443&8.872&6.936&-&\\
      43&&&18 38 56.4&-05 57 52.2&10.621&8.118&6.947&08/13/06&\\
      44&&&18 38 54.4&-05 59 53.1&8.325&7.359&7.077&-&\\
      45&&&18 39 40.1&-06 01 51.6&9.749&7.981&7.085&-&04/20/06\\
      46&&&18 39 26.8&-05 56 15.8&9.994&8.203&7.086&08/12/06&04/18/06\\
      47&&&18 39 32.5&-06 02 21.2&9.102&7.741&7.144&-&04/20/06\\
      48&&&18 39 18.6&-06 07 13.7&11.873&8.841&7.242&-&\\
      49&&&18 39 05.6&-06 04 26.6&14.228&9.919&7.324&08/12/06&04/17/06\\
      50&&&18 39 02.1&-06 02 34.1&9.676&8.204&7.394&-&\\
      51&&&18 39 20.4&-05 56 07.1&12.978&9.400&7.406&-&\\
      52&&&18 39 23.4&-06 02 15.9&11.274&8.763&7.419&05/05/06&\\
      53&&&18 39 14.5&-05 56 15.9&9.463&8.011&7.420&-&\\
      54&&&18 39 11.8&-05 58 33.7&9.746&8.169&7.462&-&\\
      55&&&18 39 12.1&-05 59 01.8&10.419&8.396&7.471&-&\\
      56&&&18 39 41.0&-05 59 10.8&10.521&8.399&7.479&-&\\
      57&&&18 39 03.9&-05 54 32.5&12.332&9.240&7.518&-&\\
      58&&&18 39 00.8&-05 59 08.2&12.261&9.086&7.522&08/13/06&\\
      59&&&18 38 50.1&-05 59 25.6&9.285&8.109&7.635&-&\\
      60&&&18 39 18.1&-06 03 08.3&10.756&8.721&7.679&08/12/06&\\
      61&&&18 39 11.8&-06 03 15.3&10.972&8.721&7.685&08/12/06&04/17/06\\
      62&&&18 39 28.6&-05 59 50.2&11.147&8.934&7.692&-&\\
      63&&&18 39 30.1&-05 57 26.9&10.799&8.732&7.717&-&04/18/06\\
      64&&&18 39 28.5&-06 06 43.6&9.246&8.148&7.749&-&\\
      65&&&18 39 06.3&-06 05 22.2&12.850&9.520&7.767&-&04/17/06\\
      \hline
    \end{tabular}
\end{table}
\addtocounter{table}{-1}
\begin{table}[hx]
    \caption{Cont.}
    \begin{tabular}{lccccccccc}
\hline\hline
      ID&S90&N01&RA&Dec&J&H&$K_{s}$& \multicolumn{2}{c}{Obs Date}\\
      & & &\multicolumn{2}{c}{(J2000)} & & & & {\it NIRSPEC} & {\it IRMOS}\\
      \hline
      66&&&18 39 13.4&-05 55 18.0&10.927&8.787&7.768&-&\\
      67&&&18 38 54.7&-05 56 55.2&13.924&9.788&7.791&-&\\
      68&&&18 39 04.5&-06 06 13.4&11.298&8.904&7.806&-&\\
      69&&&18 38 54.8&-06 04 45.3&13.447&9.676&7.886&-&\\
      70&&&18 39 25.4&-06 07 17.6&9.542&8.370&7.896&-&\\
      71&&&18 38 46.8&-06 01 23.8&12.651&9.446&7.903&-&\\
      72&&&18 39 16.2&-06 03 07.2&12.196&9.506&7.920&08/12/06&\\
      \hline
    \end{tabular}
\end{table}

\subsection{{\it IRMOS} observations}
The Infra-red Multi-Object Spectrograph \citep[{{\it IRMOS},
}][]{MacKenty03}, uses a micro-mirror array of $\sim$10$^{6}$
elements. Synthetic `slits' can be defined by switching selected
mirrors of the array into the `on' position. We split the cluster up
into sub-fields, and using pre-imaging data defined a series of slits
at the positions of stars in the current field. We interweaved science
frames with `all-off' frames, to make an accurate measure of the dark
current and instrumental background, and took flat-field observations
for each MOS configuration with a continuum lamp. We observed the
A~star HD~44612 as a telluric standard.

\subsubsection{IRMOS Data reduction}
For each science frame, we subtracted the corresponding dark frame and
extracted sub-frames containing each of the spectra. Each sub-frame
was divided through by its associated flat-field in the continuum-lamp
exposures, to correct for pixel-to-pixel variations in
sensitivity. Sky-subtraction was done by interpolation of the regions
either side of the star, and the spectra were optimally-extracted by
weighting the pixels according to the strength of the signal above the
sky background. Wavelength calibration was done using the sky emission
lines and telluric absorption features. The Br $\gamma$ absorption was
removed from the telluric standard via linear interpolation of the
continuum, and the resulting spectrum used to correct for telluric
absorption.

\subsection{{\it NIRSPEC }observations}
We used the instrument in high-resolution mode, with the NIRSPEC-7
filter, in conjunction with the 0.576\arcsec\ $\times$ 24\arcsec\
slit. The dispersion angle was set to 62.53\deg, with cross-dispersion
angle set to 35.53\deg. This gave us a spectral resolution of
$\sim$17,000 in the wavelength range of 1.9-2.4\microns.

We integrated on each star for 20s in each of two nod-positions along
the slit. In addition to the cluster stars, we also observed HD
171305, a B0~V star, as a telluric standard on each night. Flat-fields
were taken with a continuum lamp. For wavelength calibration purposes,
arc frames were taken with Ar, Ne, Xe and Kr lamps to provide as many
template lines as possible in the narrow wavelength range of each
spectral order. To fully sample the wavelength regions between the
spectral lines, we also observed the continuum lamp through an etalon
filter.

\subsubsection{ NIRSPEC Data reduction}
Removal of sky emission, dark current and bias offset was done by
subtracting nod-pairs, and images were flat-fielded with the
continuum-lamp exposures. Correction for the warping of the order
images on the detector ({\it rectification}) was done following the
method described in detail by \citet{Figer03}. An outline of the
method is presented below. 

The 2-D transformation matrix which corrects for the warping of the
orders is known as the {\it rectification matrix}, and must compensate
for the warping in both the spatial and spectral
directions. Correction in the spatial direction was found from
polynomial fitting to the two offset star traces in a nod-pair. In the
spectral direction, we first found an initial wavelength solution by
fitting a first-order polynomial to the arc lines in each order. This
solution was then applied to the etalon frames, giving an estimate of
the etalon-line wavelengths, and hence of the separation of the etalon
plates. The wavelengths of the etalon lines were then recalculated,
assuming that the wavelengths are determined by the equation
$\lambda_{n} = t/2n$, where $t$ is the plate separation and
$\lambda_{n}$ is the wavelength of the $n$th order. These recalculated
wavelengths were fitted with a 3rd-order polynomial to make a
secondary estimate of the rectification matrix. This solution was
applied to the arc frame, and the residuals between the measured and
rest-frame wavelengths of the arc lines used to fine-tune the
etalon-plate thickness.

After rectifying the data, spectra were extracted by summing the
pixels across the star trace in each channel. Relative shifts in
wavelength of $\la$4\kms, due to the star not being in the exact
centre of the slit, were corrected for by cross-correlating with a
reference spectrum around the atmospheric CO$_{2}$ feature beginning
at $\sim$ 1.95\microns. The mean shift was taken to be the
slit-centre, and the spectra of all orders shifted accordingly.

The final absolute wavelength solution, applied to all {\it NIRSPEC}
data, was accurate to better than $\pm$4\kms\ across all orders, based
on the residuals of the arc line wavelengths. The {\it internal}
wavelength error between spectra, measured from the 1.95\microns\
CO$_{2}$ feature, was less than $\pm$1\kms.

Cosmic ray hits and bad pixels were corrected for by taking the ratio
of individual exposures of the same object, and identifiying pixels
outside 5$\sigma$ of the residual spectrum. Cosmic hits were replaced
with the median value of the three neighbouring pixels either side. We
removed the H and He~{\sc i} absorption features of the telluric
standard via linear interpolation either side of the line. The
atmospheric absorption features in the science frames were then
removed by dividing through by the telluric standard. Finally, the
spectra were normalised by dividing through by the median continuum
value. From featureless continuum regions in the final spectra, we
estimate the signal-to-noise to be better than 100 for all spectra.

\subsection{Data analysis} \label{sec:anal}
\subsubsection{Temperature estimation}
We estimate the temperatures and spectral types of the stars
empirically from the equivalent width of the CO bandhead feature,
$EW_{\rm CO}$. We calibrate this method with template spectra of Red
Giants and Red Supergiants taken from \citet{K-H86}, \citet{W-H96} and
\citet{W-H97}. To do this, we define a measurement region of
2.294-2.304\microns\ in the rest frame. The definition of a robust
local continuum region is problematic, due to the dense molecular and
atomic absorption lines in this region of the spectrum. Our continuum
measurements were made from the median-average of the
2.288--2.293\microns\ region. We estimated an uncertainty on the
$EW_{\rm CO}$ measurement by making small adjustments in the
definition of the continuum region and checking the repeatibility. We
found that the measurements were stable to around 1\AA, or $\sim$5\%.

\begin{figure}[t]
  \centering
  \includegraphics[width=12cm,bb=45 15 631 430]{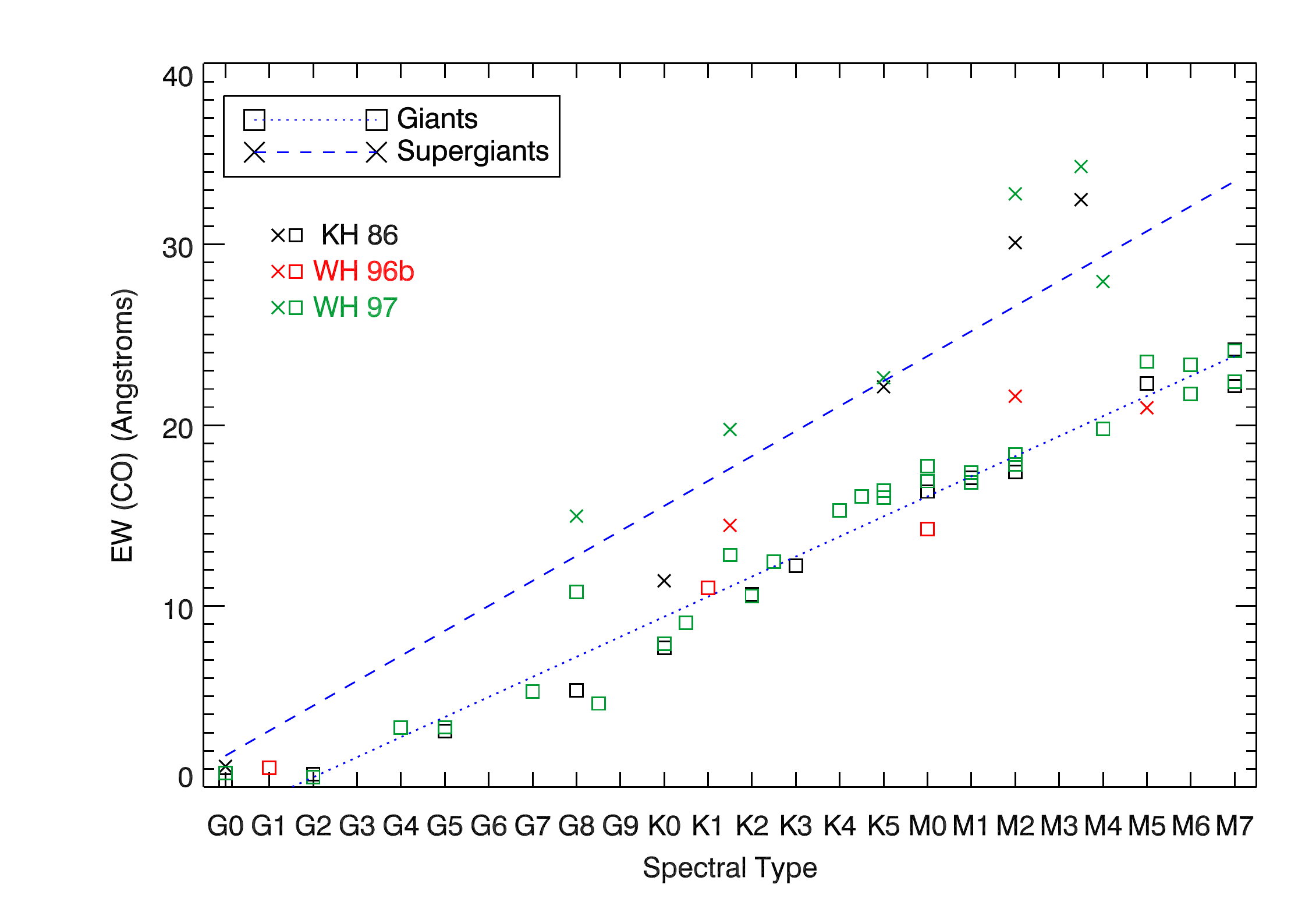}
  \caption{Relationship between spectral-type and the equivalent width
  of the CO bandhead feature, deduced from template spectra of
  \citet{K-H86}, \citet{W-H96} and \citet{W-H97}. Giants are plotted
  as squares, supergiants as crosses. The dotted and dashed lines show
  linear fits through the data. These fits were used to determine the
  spectral types of the target stars.}
  \label{fig:tempfit}
\end{figure}

Figure \ref{fig:tempfit} shows the correlation between spectral types
of the template stars and $EW_{\rm CO}$ as measured across the
predefined wavelength range. It can be seen that the relationship
between the two is approximately linear, while for a given spectral
type supergiants tend to show stronger CO absorption than giants. We
determined the spectral types of those stars classified as supergiants
from the linear fit to the literature data shown in Fig.\
\ref{fig:tempfit}. For the template giants the rms scatter on the fit
is $\pm$1 subtype, while the scatter is larger for the template
supergiants ($\pm$2 subtypes). This is the uncertainty we adopt
throughout the rest of this paper. In converting spectral type to
effective temperature, we use the temperature scale recently rederived
by \citet{Levesque05}.


\subsubsection{Radial velocity measurements}
Following the method described in detail by \citet{Figer03}, accurate
radial velocities were measured by cross-correlating the complex
stellar CO bandhead feature beginning at $\sim$2.293\microns\ with
that of Arcturus, from the spectrum of \citet{W-H96arct} shifted to
the zero local standard of rest velocity. We experimented with using
different wavelength regions for the cross-correlation, such as
including/excluding the sharp edge at the blue-edge, to test the
robustness of the measurement. We found that our velocity measurements
are stable to within $\pm$1\kms\ regardless of the wavelength range
used, therefore the absolute uncertainty of any velocity measurement
is dominated by the $\pm$4\kms\ error in the wavelength solution. 

\citet{Figer03} concluded that this method, when applied to Red
Giants, introduced a systematic error in the measured radial velocity
as a function of $EW_{\rm CO}$, and accounted for $\pm$2\kms\ across
the full range of observed $EW_{\rm CO}$. In order to assess any
impact this may have on our data, we applied the method to the
high-resolution template spectra of RSGs presented in
\citet{W-H97}. As these spectra are few, it was inconclusive as to
whether this systematic trend of measured $v_{\rm rad}$ with $EW_{\rm
CO}$ is present in our data, but if so it would appear to be very
small ($\pm$1\kms). It does not therefore contribute significantly to
the absolute uncertainty of individual radial velocity
measurements. It may however have an impact on the internal error
between measurements, and become important when determining the
cluster's virial mass. We will make a more comprehensive discussion of
the effect of including this uncertainty in Sect.\ \ref{sec:virial}.

\section{Results} \label{sec:results}

\begin{figure}[t]
  \centering
  \includegraphics[width=8cm,bb=60 10 553 420]{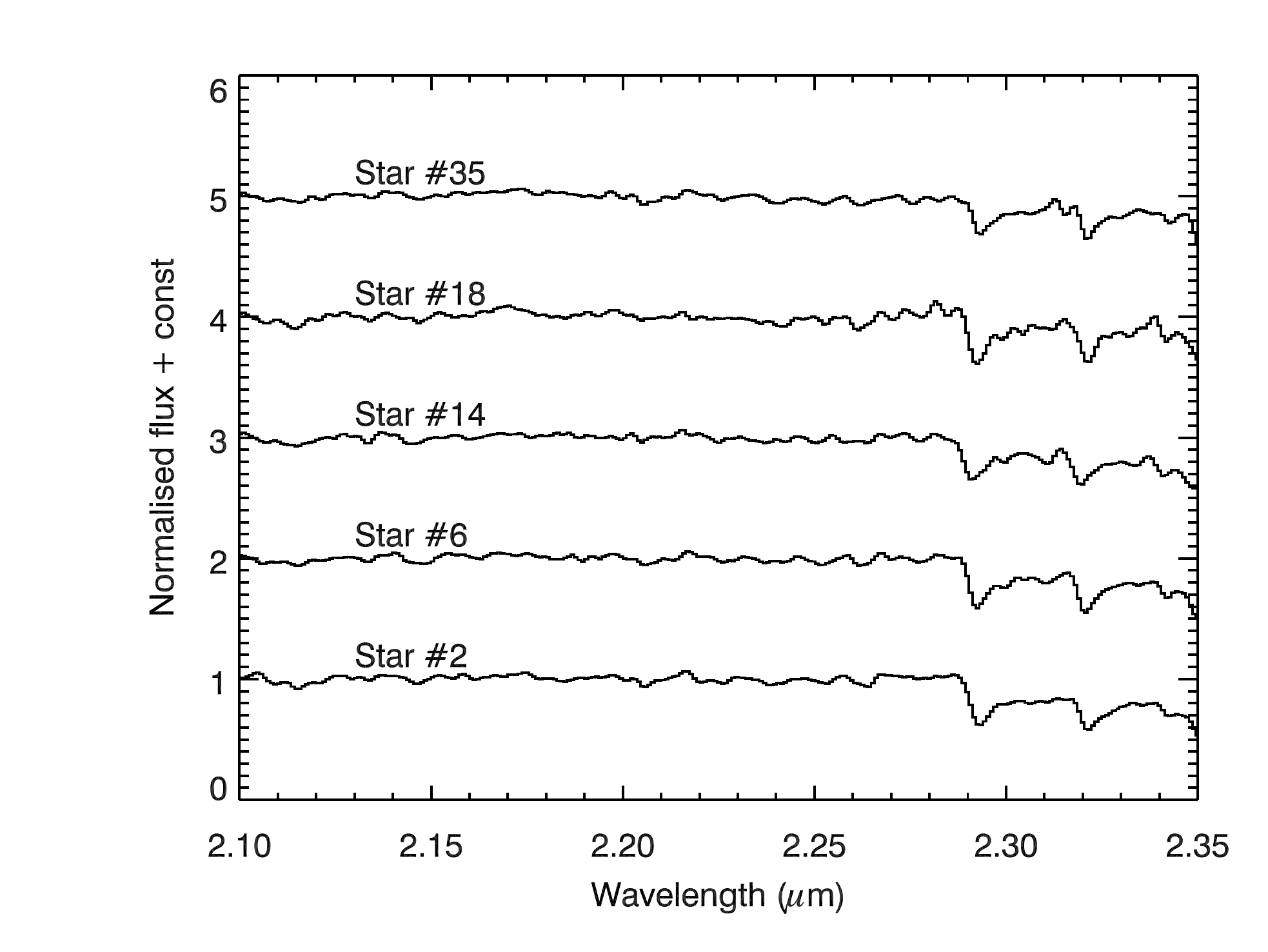}
  \includegraphics[width=8cm,bb=60 10 553 420]{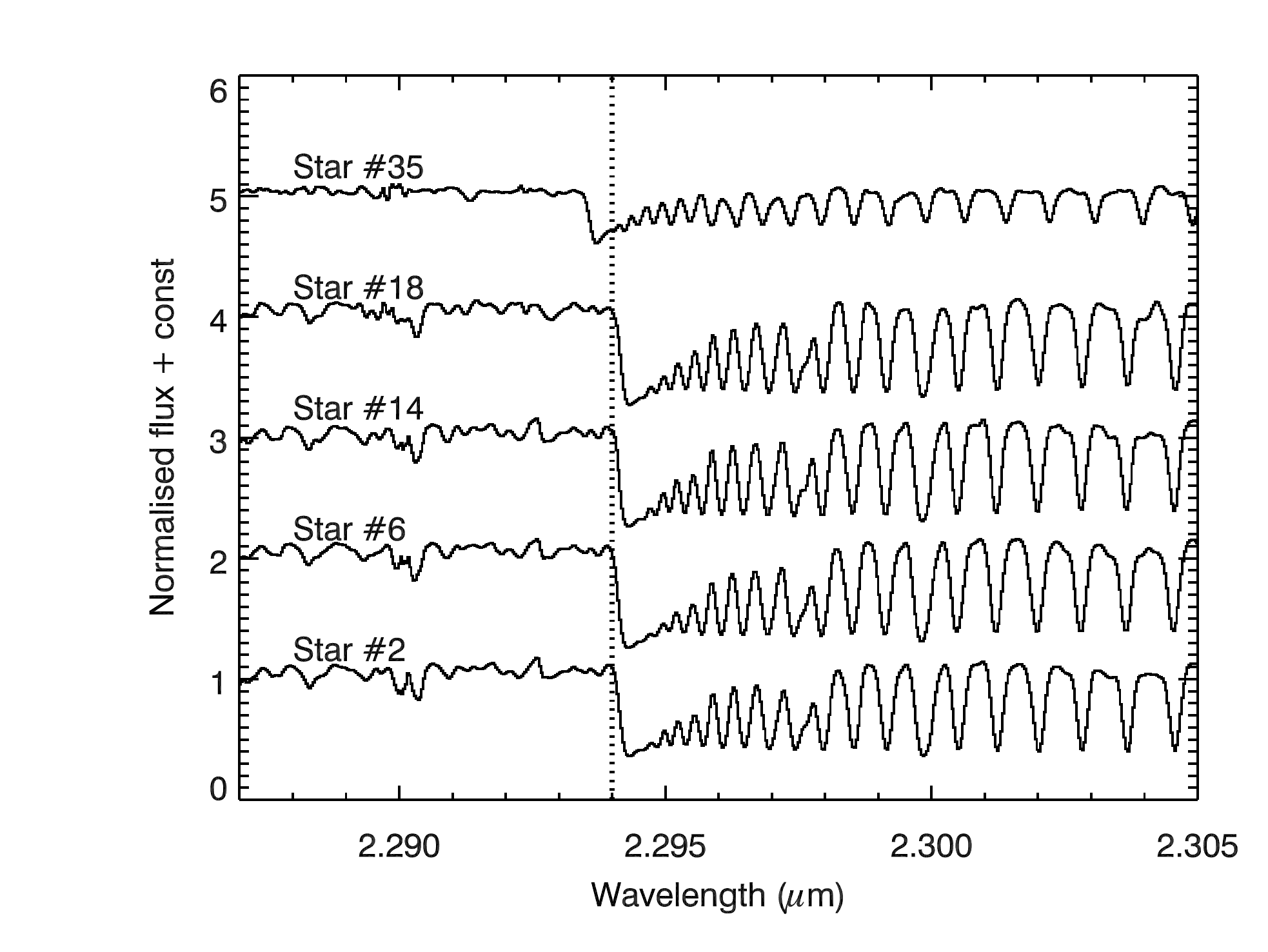}
  \caption{Examples of the spectra. {\it Left:} the {\it IRMOS} data,
  clearly showing the CO bandhead absorption -- indicative of late
  spectral-types. {\it Right:} the {\it NIRSPEC} follow-up data, which
  highlights the radial velocity differences between Star 35 and the rest.}
  \label{fig:spectra}
\end{figure}

\subsection{Spectra}
Examples of the NIR spectra are shown in Fig.\ \ref{fig:spectra}. From
the {\it IRMOS} data (Fig.\ \ref{fig:spectra}, {\it left}), the CO
bandhead absorption can clearly be seen in the spectra, implying late
spectra types for the stars. Also, no stars show the `vignetting' at
the edges of the $K$-band, indicative of the H$_{2}$O absorption often
seen in Red Giants \citep[see e.g.\ atlas of][]{K-H86}, though we note
that care should be taken using this selection criterion in
uncalibrated spectra. We therefore identify these stars as {\it
candidate} RSGs. We found stars displaying these features down to
$K_{S} \la $7.5.

%

The follow-up {\it NIRSPEC} data of these stars are shown in Fig.\
\ref{fig:spectra} ({\it right}), and show the blue-edge of the feature
in greater detail. Here, the difference between Star 35 and the others
shown in the plot is obvious -- it can be seen from the blue-edge of
the CO bandhead feature (indicated by the dotted line) Stars 2, 6, 14
and 18 have very similar velocities. Star 35 however, has a notable
velocity-shift with respect to the others. In the following section,
we use the radial velocity and CO equivalent width measurements to
argue that, of the stars observed, a total of 26 can be readily
identified as being part of a physical association of RSGs.

\begin{table}[p]
  \centering
  \caption{Derived data of the cluster stars. (1): Star ID, in order
    of ascending brightness in $K_{S}$; (2): radial velocity --
    absolute uncertainty is $\pm$4\kms, internal error is $\pm$1\kms;
    (3 \& 4): effective temperature and spectral type, accurate to 2
    subtypes \citep[temperatures taken from][]{Levesque05}; (5):
    $K$-band extinction; (6) absolute magnitude; (7): Luminosity.  }
  \begin{tabular}{ccccccc}
    \hline \hline
    (1) & (2) & (3) & (4) & (5) & (6) & (7) \\
    ID & $V_{\rm LSR}$ (\kms) & $T_{\rm eff}$~(K) & Spec Type & $A_{K_{S}}$ & $M_{K}$ & log($L_{bol}$/\lsun) \\    
    \hline
 2 & 111.1 & 3605$\pm$147 & M3 & 1.39$\pm$0.06 & -11.12$^{+0.33}_{-0.63}$ &   5.22$^{+0.25}_{-0.13}$ \\
 3 & 110.5 & 3535$\pm$130 & M4 & 1.34$\pm$0.07 & -10.72$^{+0.33}_{-0.63}$ &   5.04$^{+0.25}_{-0.13}$ \\
 5 & 113.3 & 3535$\pm$130 & M4 & 1.99$\pm$0.07 & -11.02$^{+0.33}_{-0.63}$ &   5.16$^{+0.25}_{-0.13}$ \\
 6 & 107.1 & 3450$\pm$100 & M5 & 1.17$\pm$0.08 &  -9.95$^{+0.32}_{-0.63}$ &   4.70$^{+0.25}_{-0.13}$ \\
 8 & 104.1 & 3840$\pm$135 & K5 & 1.45$\pm$0.09 & -10.23$^{+0.32}_{-0.63}$ &   4.94$^{+0.25}_{-0.13}$ \\
 9 & 111.8 & 3450$\pm$100 & M5 & 1.64$\pm$0.08 & -10.28$^{+0.32}_{-0.63}$ &   4.84$^{+0.25}_{-0.13}$ \\
10 & 112.1 & 3450$\pm$100 & M5 & 1.42$\pm$0.08 & -10.03$^{+0.32}_{-0.63}$ &   4.73$^{+0.25}_{-0.13}$ \\
11 & 110.1 & 3535$\pm$130 & M4 & 1.42$\pm$0.08 & -10.08$^{+0.33}_{-0.63}$ &   4.78$^{+0.25}_{-0.13}$ \\
13 & 111.5 & 3535$\pm$130 & M4 & 1.42$\pm$0.07 &  -9.85$^{+0.33}_{-0.63}$ &   4.69$^{+0.25}_{-0.13}$ \\
14 & 108.2 & 3605$\pm$147 & M3 & 1.39$\pm$0.06 &  -9.77$^{+0.33}_{-0.63}$ &   4.68$^{+0.25}_{-0.13}$ \\
15 & 112.2 & 3660$\pm$127 & M2 & 1.27$\pm$0.06 &  -9.59$^{+0.32}_{-0.63}$ &   4.63$^{+0.25}_{-0.13}$ \\
16 & 108.3 & 3605$\pm$147 & M3 & 1.25$\pm$0.06 &  -9.50$^{+0.33}_{-0.63}$ &   4.57$^{+0.25}_{-0.13}$ \\
17 & 101.4 & 4015$\pm$130 & K3 & 1.77$\pm$0.05 &  -9.99$^{+0.31}_{-0.62}$ &   4.90$^{+0.25}_{-0.12}$ \\
18 & 111.2 & 3535$\pm$130 & M4 & 1.16$\pm$0.07 &  -9.36$^{+0.33}_{-0.63}$ &   4.50$^{+0.25}_{-0.13}$ \\
19 & 106.7 & 3605$\pm$147 & M3 & 1.12$\pm$0.06 &  -9.17$^{+0.33}_{-0.63}$ &   4.44$^{+0.25}_{-0.13}$ \\
20 & 109.9 & 3660$\pm$127 & M2 & 1.30$\pm$0.06 &  -9.32$^{+0.32}_{-0.63}$ &   4.52$^{+0.25}_{-0.13}$ \\
21 & 107.4 & 3660$\pm$127 & M2 & 1.79$\pm$0.06 &  -9.81$^{+0.32}_{-0.63}$ &   4.71$^{+0.25}_{-0.13}$ \\
23 & 119.3 & 3535$\pm$130 & M4 & 2.29$\pm$0.07 & -10.35$^{+0.33}_{-0.63}$ &   4.89$^{+0.25}_{-0.13}$ \\
26 & 107.6 & 3605$\pm$147 & M3 & 1.31$\pm$0.06 &  -9.16$^{+0.33}_{-0.63}$ &   4.44$^{+0.25}_{-0.13}$ \\
27 & 112.8 & 3660$\pm$127 & M2 & 1.45$\pm$0.06 &  -9.19$^{+0.32}_{-0.63}$ &   4.47$^{+0.25}_{-0.13}$ \\
29 & 104.9 & 3790$\pm$130 & M0 & 1.14$\pm$0.08 &  -8.86$^{+0.33}_{-0.63}$ &   4.38$^{+0.25}_{-0.13}$ \\
30 & 107.9 & 3745$\pm$117 & M1 & 1.15$\pm$0.09 &  -8.82$^{+0.32}_{-0.63}$ &   4.35$^{+0.25}_{-0.13}$ \\
31 & 105.2 & 3745$\pm$117 & M1 & 1.62$\pm$0.09 &  -9.24$^{+0.32}_{-0.63}$ &   4.51$^{+0.25}_{-0.13}$ \\
49 & 109.4 & 3920$\pm$112 & K4 & 4.58$\pm$0.22 & -11.30$^{+0.38}_{-0.66}$ &   5.39$^{+0.26}_{-0.15}$ \\
52 & 111.2 & 3790$\pm$130 & M0 & 2.24$\pm$0.08 &  -8.72$^{+0.33}_{-0.63}$ &   4.32$^{+0.25}_{-0.13}$ \\
72 & 107.3 & 3790$\pm$130 & M0 & 2.55$\pm$0.16 &  -8.62$^{+0.36}_{-0.64}$ &   4.28$^{+0.26}_{-0.14}$ \\
    \hline
  \end{tabular}
  \label{tab:clresults}
\end{table}

\subsection{Supergiants vs. foreground stars} \label{sec:gsg}

\begin{figure}[t]
  \centering
  \includegraphics[width=\columnwidth,bb=32 10 992 510]{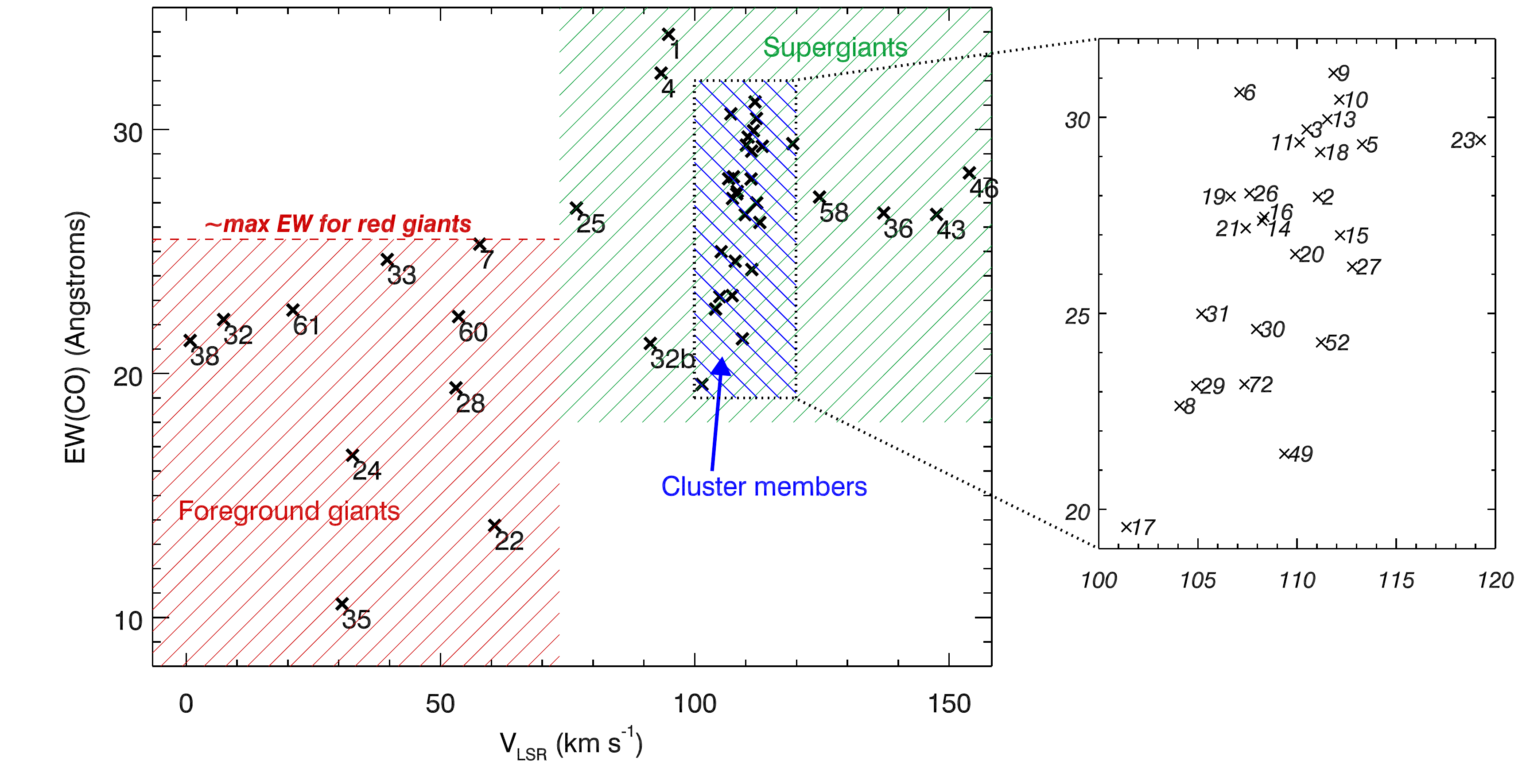}
  \caption{Plot of CO equivalent width versus $v_{\rm LSR}$ for all
  stars observed with {\it NIRSPEC}. Stars in the cross-hatched region
  have very similar velocities, and are likely to be part of the same
  cluster. Stars in the bottom-left are probably foreground stars due
  to their low \EW\ and much lower velocities. }
  \label{fig:v_ew}
\end{figure}

\begin{figure}[t]
  \centering
  \includegraphics[width=12cm,bb=40 30 695 630]{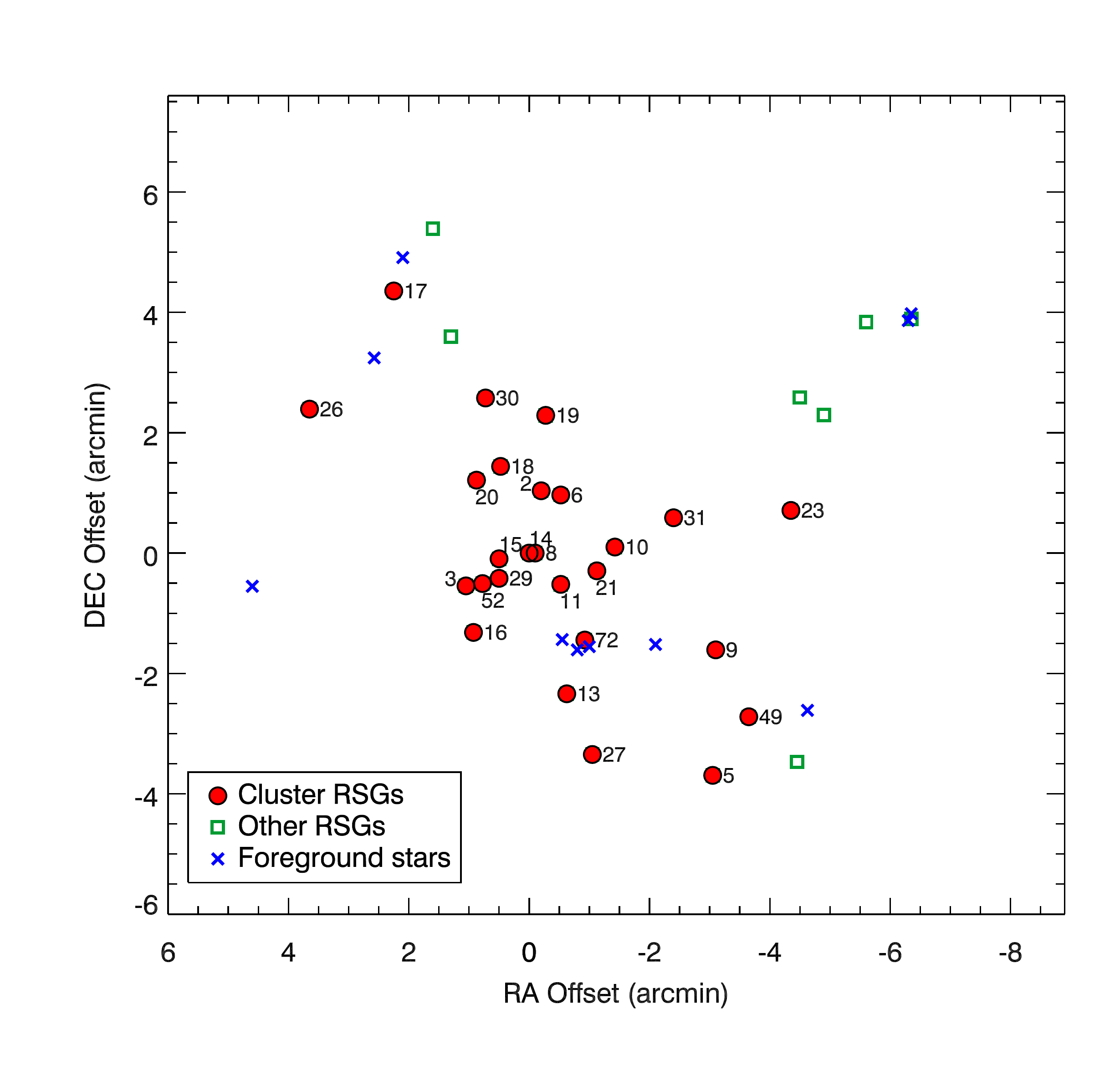}
  \caption{Illustration of the location of the stars belonging to the
  cluster, and foreground/background objects.  }
  \label{fig:clusterfinder}
\end{figure}

The observed stars in this field separate into three categories:
physically-associated cluster members, unrelated foreground/background
RSGs, and foreground M giants and dwarfs. Here we use the
observational data from the high-resolution spectroscopy to determine
which stars lie in which of these categories.

Figure \ref{fig:v_ew} shows a plot of CO bandhead equivalent width
($W_{\lambda}$) against radial velocity. Marked on the plot is the
maximum $EW_{\rm CO}$ of the CO bandhead feature observed in M
giants. Any stars above this line are therefore likely to be
supergiants. The plot shows that there are no stars with high-$EW_{\rm
CO}$ at low radial velocities, just as there are no stars with
low-$EW_{\rm CO}$ at high-$v_{\rm rad}$. We thus identify those
stars in the lower-left of the plot (red crosshatched region) as
foreground stars; while those stars in the upper-right, with higher
$v_{\rm rad}$ and $EW_{\rm CO}$, are identified as more-distant
supergiants.

Notice that within this subsample of `supergiants', there is a tight
grouping of many stars with radial velocities 100--120\kms. We
identify the 26 stars within this grouping (crosshatched region) as
being part of a physical association. A cluster member with a velocity
outside this range would imply a large runaway speed, and would be
unlikely to remain in the field-of-view for a likely cluster age of
$\sim$10Myr (see Sect.\ref{sec:age}). The supergiants ouside this
region are therefore probably unrelated objects in the same
line-of-sight along the base of the Scutum-Crux arm. It will be shown
later that these objects are typically more reddened, supporting this
conclusion (see Sect. \ref{sec:extinct}). There is of course the
possibility that the sample of 26 `cluster' stars is contaminated by
other RSGs along the line-of-sight with peculiar velocities, making
them appear to be part of the cluster. This number is difficult to
quantify, but from the velocity-spread of the `unrelated' supergiants
it would seem unlikely that there more than one or two interlopers.

From these selection criteria, we determine that the cluster contains
26 RSGs -- the largest associated population of RSGs discovered to
date and almost twice as many as in the nearby RSGC1 described in
FMR06. We also identify a further 8 RSG candidates along the
line-of-sight to RSGC2, based on the stars' high CO equivalent
widths. The location of these stars within the field of the cluster is
illustrated in Fig.~\ref{fig:clusterfinder}. 


Aside from Stars 12 \& 34, our high-resolution observations are
complete down to $K_{S} < $6.6. Of the 33 stars brighter than this
threshold, 23 are determined to be cluster members. Of the stars we
sampled fainter than this threshold, only 3 of 11 stars were found to
belong to the cluster. Hence the cluster RSGs tend to be among the
brighter stars in the field, and it is unlikely that a significant
number of cluster RSGs (more than one or two) were missed in our
survey of the cluster.

\subsection{Spectral types} \label{sec:spectypes}
\begin{figure}[t]
  \centering 
  \includegraphics[width=12cm,bb=60 15 620 490]{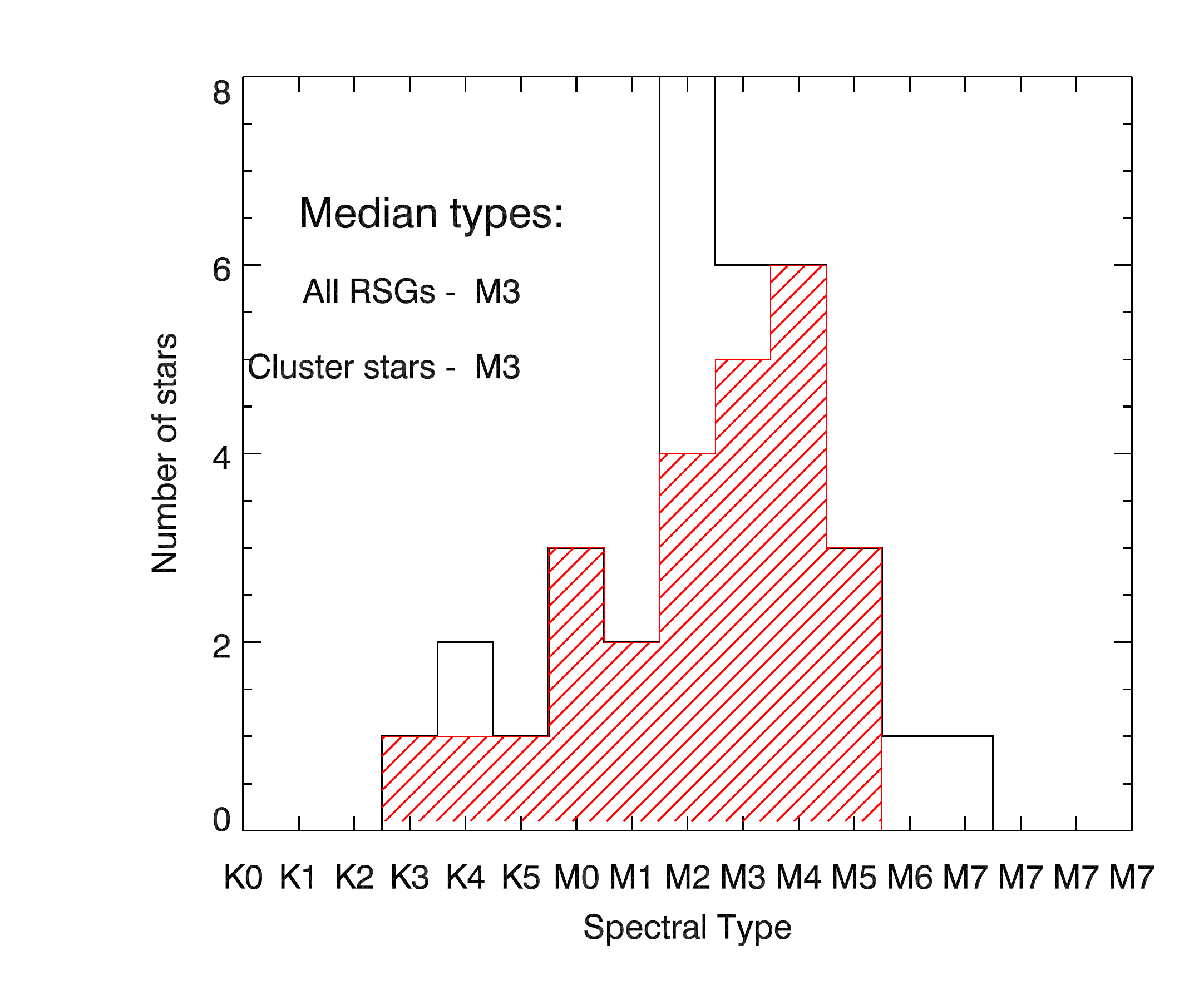} 
  \caption{Histogram of the spectral types of the
  stars identified as supergiants, and the subsample determined to be
  `cluster members' (cross-hatched region). The median type of both
  subsamples (M3) is consistent with the nearby RSG cluster RSGC1 (M3,
  FMR06) and the rest of the Galaxy \citep[M2, ][]{Elias85}.}
  \label{fig:sp_hist}
\end{figure}

We take the stars classified as supergiants and determine their
spectral types based on the relation to CO-bandhead equivalent width
derived in Sect. \ref{sec:anal}. The spectral types are listed in
Table \ref{tab:clresults}, and are plotted in a histogram in
Fig.\,\ref{fig:sp_hist}. We plot both the entire sample of
supergiants, and the subsample of cluster members (cross-hatched). The
median spectral-type for both samples is M3~{\sc i}, which is in
agreement with the average spectral type in the Galaxy (M2) and that
of the nearby RSGC1 cluster (M3) \citep[][FMR06]{Elias85}.

The median spectral-type of RSGs is thought to be linked to chemical
abundance. For example, RSG distribution in the low-metallicity
environments of the Magallic Clouds is significantly different to that
of the Galaxy. Median spectral-types of K5, M1, and M2 have been found
from studies of the SMC, LMC and the Galaxy respectively
\citep{Humphreys79,Elias85,M-O03}. This difference is thought to arise
either from the metallicity-dependent opacity of the stellar envelope
\citep{Elias85}, or the abundance-sensitive strengths of diagnostic
molecular absorption features, e.g.\ TiO bands \citep{M-O03}. Either
way, that the median spectral-types of the clusters RSGC1 and 2 agree
with i) each other, and ii) those in the rest of the Galaxy, suggests
that these objects have roughly similar metallicities to the Galactic
average.

\subsection{Cluster distance} \label{sec:dist}
The mean radial velocity of the cluster subsample is $\bar{v}_{LSR} =
109.3\pm 0.7$\kms, with the uncertainty taken from Poisson statistics
of the 26 stars. The cluster radial velocity is constrained extremely
well by the large number of measurements, and the uncertainty in this
value is dominated by the precision of the wavelength solution, $\pm
4$\kms. 

\begin{figure}[t]
  \centering
  \includegraphics[width=12cm,bb=30 15 620 490]{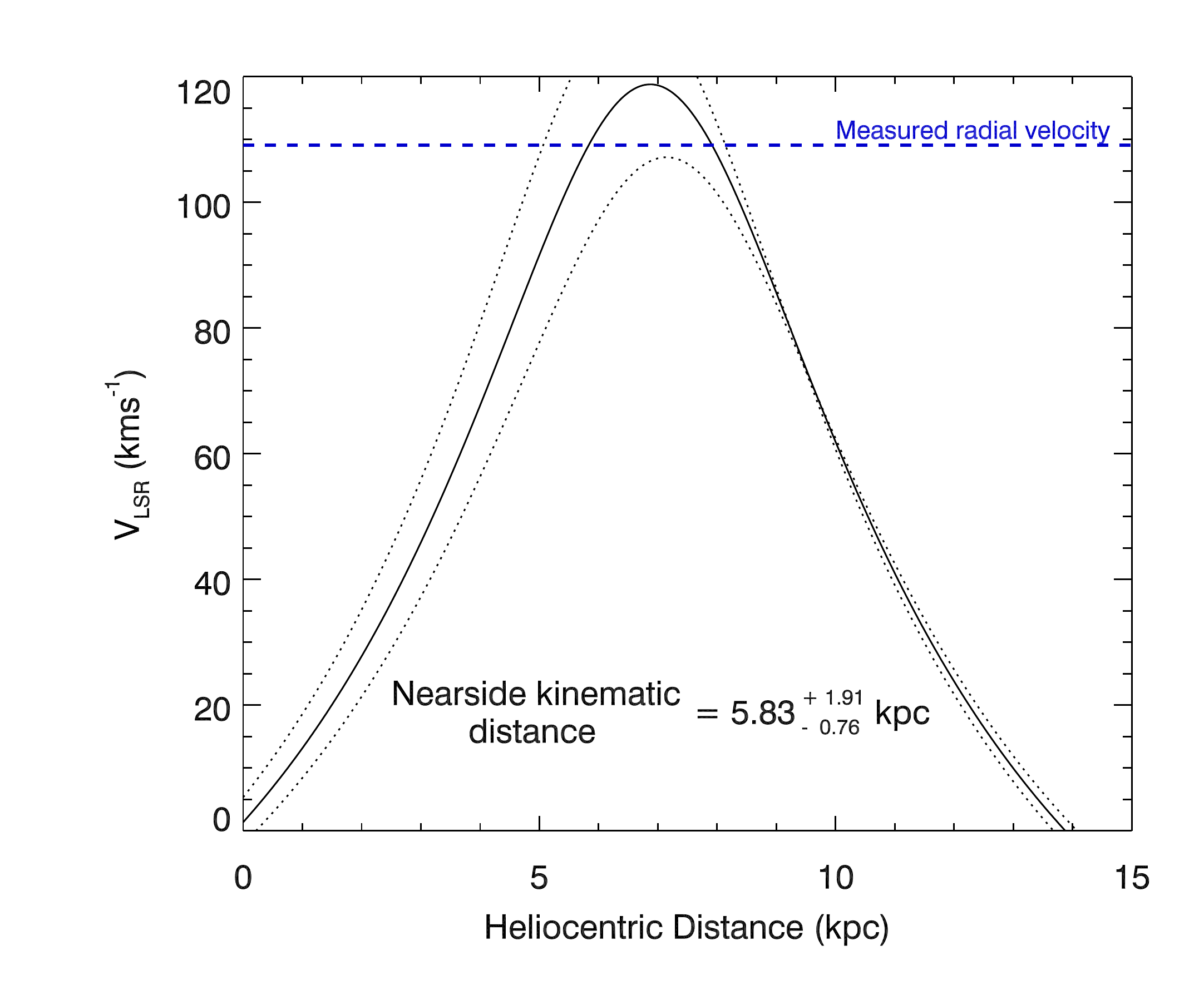}
  \caption{Galactic rotation curve in the direction of RSGC2, using
  the latest measurements collated by \citet{K-D07} (solid-line). The
  dashed-line represents the mean radial velocity of the cluster
  stars, implying the kinematic distance marked on the
  plot. Dotted-lines illustrate the errors in this distance implied by
  uncertainties in the solar rotational velocity and the distance to
  the Galactic centre. }
  \label{fig:grot}
\end{figure}

In converting this radial velocity into a kinematic distance to the
cluster, we are limited by the uncertainties in the Galactic rotation
curve. We use the most contemporary measurements of the Galactic
centre distance and solar rotational velocity as compiled by
\citet[in press]{K-D07}. In determining the
distance to Wd~1, these authors used the Galactic centre distance
$D_{\rm GAL} = $7.6$\pm$0.3~kpc, as determined by
\citet{Eisenhauer05}, and the solar rotation velocity of
$\Theta_{\odot} = $214$\pm$7~\kms, averaged from measurements by
\citet{R-B04} and \citet{F-W97}. We use these values to construct the
Galactic rotation curve in the direction of RSGC2 shown in Fig.\
\ref{fig:grot}. 

From comparison with the cluster's radial velocity, we derive a
kinematic distance of $5.83^{+1.91}_{-0.78}$~kpc. The uncertainties
are determined from the minimal and maximal nearside distance from the
errors in $D_{\rm GAL}$ and $\Theta_{\odot}$, and are rather large due
to the location of the cluster close to the tangential point of the
Galactic arm (see Fig.\ \ref{fig:grot}).

This distance estimate is considerably closer than the $\sim$30kpc
quoted in S90. This greater distance was determined by assuming that
the cluster stars were M supergiants, then calculating the distance
modulus based on their typical absolute magnitudes. However, S90 did
not take into account interstellar extinction, which we determine to
be $A_{V} = 13.1$ assuming the interstellar extinction law of
\citet{R-L85} (see Sect.~\ref{sec:extinct}). This explains S90's large
distance overestimate.

In a separate study, \citet{Nakaya01} derived a much closer distance
estimate -- they isolated a sample of stars which appeared to be
reddened in their $I$ vs.~$R-I$ colour-magnitude diagram, and assumed
these were early-type stars belonging to the cluster. They then
dereddened these stars to the intrinsic colours of A0 stars, deriving
an extinction of $A_{V} = 11.4$. Upon fitting the `A0' stars and the
S90 `red' stars with a model isochrone, they finally arrived at a
distance of 1.5kpc. In analysis of similar data, \citet{Ortolani02}
derived a distance of 6kpc, by adding the constraint that the cluster
containing so many RSGs cannot be older than $\sim$20Myr.

While Ortolani et al.'s distance estimate is comparable to ours, these
studies highlight the problematic nature of inferring cluster
properties from photometry alone. Our method of determining the
distance to the cluster is much more direct and relies less on
assumptions of cluster membership and spectral types. From the radial
velocity data, the grouping of so many stars with a velocity rms of
$\pm3.5$\kms\ is strong evidence that these stars are associated. Even
allowing for a peculiar cluster velocity of $\pm$20\kms\ from the
Galactic rotation curve, due to e.g.\ the cluster's proximity to the
Galactic bulge, this would still only imply an extra uncertainty of
$\pm$1kpc.

\begin{figure}[t]
  \centering
  \includegraphics[width=12cm,clip,bb=60 15 600 480]{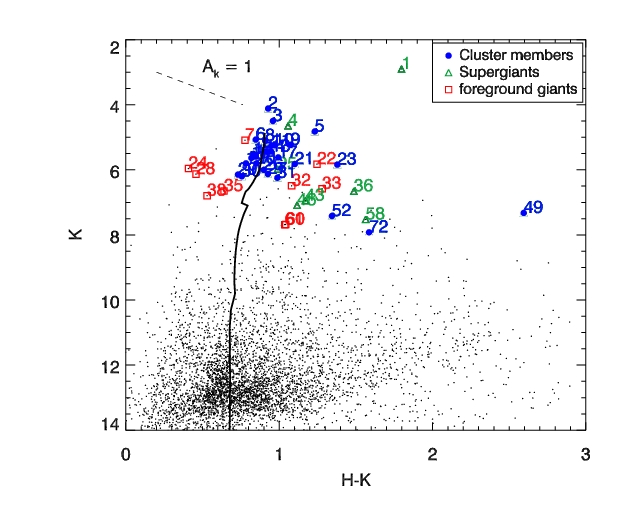}
  \caption{Colour -- magnitude diagram for all stars in the {\it
  2MASS} point-source catalogue within 7\arcmin\ of the cluster center
  ({\it dots}). The stars observed in this paper are plotted with
  their ID\# as indicated in the legend. The solid line represents a
  12Myr isochrone from Geneva models with solar metallicity and
  canonical mass-loss rates, which has been reddened according to the
  interstellar extinction-law of \citet{R-L85} with $A_{K} = 1.47$. A
  reddening vector of $A_{K} = 1$ is also plotted. The majority of the
  cluster stars form a tight grouping at the top of the isochrone, the
  location of RSGs.}
  \label{fig:colmag_hk}
\end{figure}
\begin{figure}[t]
  \centering
  \includegraphics[width=12cm,clip,bb=60 15 600 480]{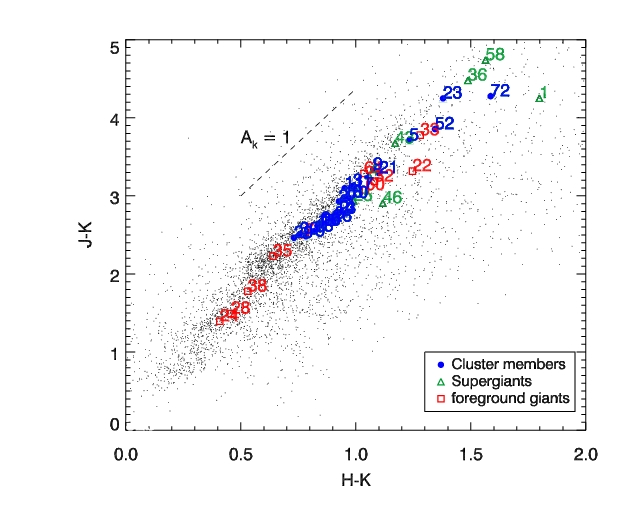}
  \caption{{\it 2MASS} colour-colour diagram of the stars plotted in
  Fig.~\ref{fig:colmag_hk}, with the same plotting symbols. A
  reddening vector according to $A_{K} = 1$ is plotted. For clarity,
  Star~49 -- which lies along the reddening vector from the cluster
  stars with an extinction $A_{K} = 4.58$ -- is not shown.}
  \label{fig:colcol}
\end{figure}

\subsection{Extinction} \label{sec:extinct}
Figure \ref{fig:colmag_hk} shows a 2MASS colour-magnitude diagram of
all stars with 7\arcmin\ of the cluster centre, which is taken to be
the position of Star 14 (see Table \ref{tab:obsdata}). The stars
classified as cluster members are plotted as circles, the unrelated
supergiants as triangles, and the foreground stars as squares. An
$A_{K} = 1$ reddening vector is plotted, and to guide the eye a 12Myr
solar metallicity isochrone is shown \citep{Schaerer93}, which has
been dereddened by $A_{K} = 1.47$ (see below).

It can be seen that the majority of the cluster stars are grouped
tightly at the top of the isochrone, with no cluster stars located
significantly to the left of the isochrone -- consistent with the
sample being uncontaminated by foreground stars. The cluster stars
significantly to the right of the isochrone all lie along a reddening
vector from this core grouping, suggesting that extra circumstellar
reddening exists for these objects. A 2MASS colour-colour diagram is
shown in Fig. \ref{fig:colcol}. Again, the majority of the cluster
stars form a tight grouping, and are located along the reddening
vector from the origin.

To determine the extinction to the cluster, we de-redden each cluster
star to the intrinsic colours appropriate for their spectral type and
a luminosity class of Iab, according to the survey of Galactic RSGs by
\citet{Elias85}. From the ($J-K_{S}$) and ($H-K_{S}$) colour excesses
of each star, we determine the extinction $A_{K_{S}}$ towards each
star using the relation in \citet{R-L85}:

\begin{equation}
A_{K_{S}} = \frac{ E_{\lambda - K_{S}} } 
{ (\lambda / \lambda_{K_{S}})^{-1.53} - 1}
\end{equation}

\noindent The uncertainty in each star's extinction is governed by the
error in its spectral-type. For $\pm$2 subtypes, then from the
variations in RSG intrinsic colours the uncertainty in the $A_{K}$
measurement of each star is about $\pm$0.06. This is consistent with
the differences we find in the extinction measurements using the two
different colour-excesses (see below).

From the measurements of all the stars, we find the median extinction
towards the cluster to be $A_{K_{S}} (J-K_{S}) = 1.462$, and
$A_{K_{S}} (H-K_{S}) = 1.424$. These two measurements are consistent
with one another, and we adopt the average of these measurements,
$A_{K_{S}} = 1.44 \pm 0.02$, and hence $A_{V} = 12.9 \pm 0.2$, to be
the extinction towards RSGC2. The measured extinction towards the
individual stars, and the associated uncertainty determined from the
error in spectral-type, is listed in Table \ref{tab:clresults}.

As mentioned in Sect.\ \ref{sec:dist}, this extinction is slightly
higher than that derived by \citet{Nakaya01}, who found $A_{V} =
11.2$. Their extinction estimate was based on the assumption that a
collection of stars with particular $R-I$ colours were cluster members
and had mean spectral-type A0. As we know the spectral types of the
stars to within a subtype, and are able to take the median of many
stars which we are confident are cluster members, we consider ours to
be a much more reliable estimate.

From Table \ref{tab:clresults}, we see that the extinction towards the
individual stars varies substantially across the field. In particular,
Star~49 is significantly more obscurred than the rest, with
$A_{K_{S}}=4.6$. In the case of this star, the extra extinction can be
readily associated with mid-IR excess, and is likely due to
circumstellar material (see Sects.\ \ref{sec:lum} and
\ref{sec:others}). For the other stars with extra extinction, such as
Stars 5, 23, 52 and 72, an association with mid-IR excess is less
obvious: \#5 and \#23 have only modest excess (see Fig.\
\ref{fig:seds}), while \#52 and \#72 are not detected in {\it
MSX}. These anomalies may arise due to a combination of factors: there
may be significant non-uniformity in the foreground extinction, indeed
it can be seen in Fig.\ \ref{fig:comp} that there is much diffuse
8\micron\ emission over the field (with \#23 appearing to be
spatially-coincident with a dark-lane); source-confusion in the {\it
MSX} images may make it impossible to detect fainter stars in crowded
regions, in particular \#52, which is dwarfed by emission from the
nearby \#6, may suffer from this effect; and finally we cannot
entirely discount that the sample of `cluster' stars is contaminated
by one or two background stars with peculiar velocities, which are
extincted by an increased column-density of interstellar material (see
also Sect.\ \ref{sec:gsg}).

\subsection{Luminosities and spectral energy distributions} \label{sec:lum}
From each stars' extinction and the kinematic distance of the cluster,
we calculate absolute magnitude of each star $M_{K_{S}}$. We then
interpolate the bolometric correction $BC_{K}$ for each star's
temperature, according to the recently re-derived values of
\citet{Levesque05}, to estimate their bolometric luminosities. These
results are listed in Table \ref{tab:clresults}.

The uncertainties in $L_{\rm Bol}$ are derived from the quadrature sum
of the errors in $A_{K}$, $BC_{K}$, and the cluster distance, $D_{\rm
cl}$. The errors in $A_{K}$ and $BC_{K}$ are governed by the precision
to which we can determine the stars' spectral types, i.e.\ $\pm$2
subtypes. While these uncertainties are small compared to that in
$D_{\rm cl}$, we can be confident that the stars are all at the {\it
same} distance. Hence, the propagation of the error in spectral-type
through to that in $L_{\rm Bol}$ will be important when investigating
the luminosity spread of the RSGs and the age of the cluster.

In addition to the {\it 2MASS} photometry, we also identify these
stars in the point-source catalogues of {\it Spitzer/GLIMPSE}
\citep{Benjamin03} and {\it MSX} \citep{Egan01}. In the {\it GLIMPSE}
catalogue, many of the stars are too bright to be included in the
high-precision version of the catalogue, and instead only appear in
the less-accurate, `more-complete' version. In the case of the {\it
MSX} data, despite the large beamsize ($\sim$18\arcsec), the cluster
is open enough to get unambiguous photometry on several of the stars.

In Fig.\ \ref{fig:seds} (Appendix A) we plot the spectral energy
distributions (SEDs) of the cluster stars. We de-redden the fluxes of
the stars according to the extinction $A_{K_{S}}$ towards each star,
in combination with the interstellar extinction law for {\it GLIMPSE}
and {\it MSX} photometry as defined in \citet{Indebetouw05} and
\citet{Messineo05} respectively. The raw photometry is plotted as
crosses, and the dereddened as filled circles.

Plotted over the photometry are black-body curves appropriate for
stars' temperatures, absolute $K$-band magnitudes and the nearside
kinematic distance to the cluster. For the majority of the stars, the
black-body curves provide good fits to the dereddened photometry up to
8\microns, and validates our empirical method of determining the
star's temperatures. Some objects, e.g.\ Stars 19, 20 and 21, appear
to be under-luminous at 4.3\microns. This can be understood as being
due to dense molecular absorption bands in this wavelength range,
which can readily be seen in spectral-type M model-atmospheres
\citep[e.g.\ ][]{Fluks94}.

\begin{figure}[p]
  \centering
  \includegraphics[width=16cm,bb=0 0 566 236]{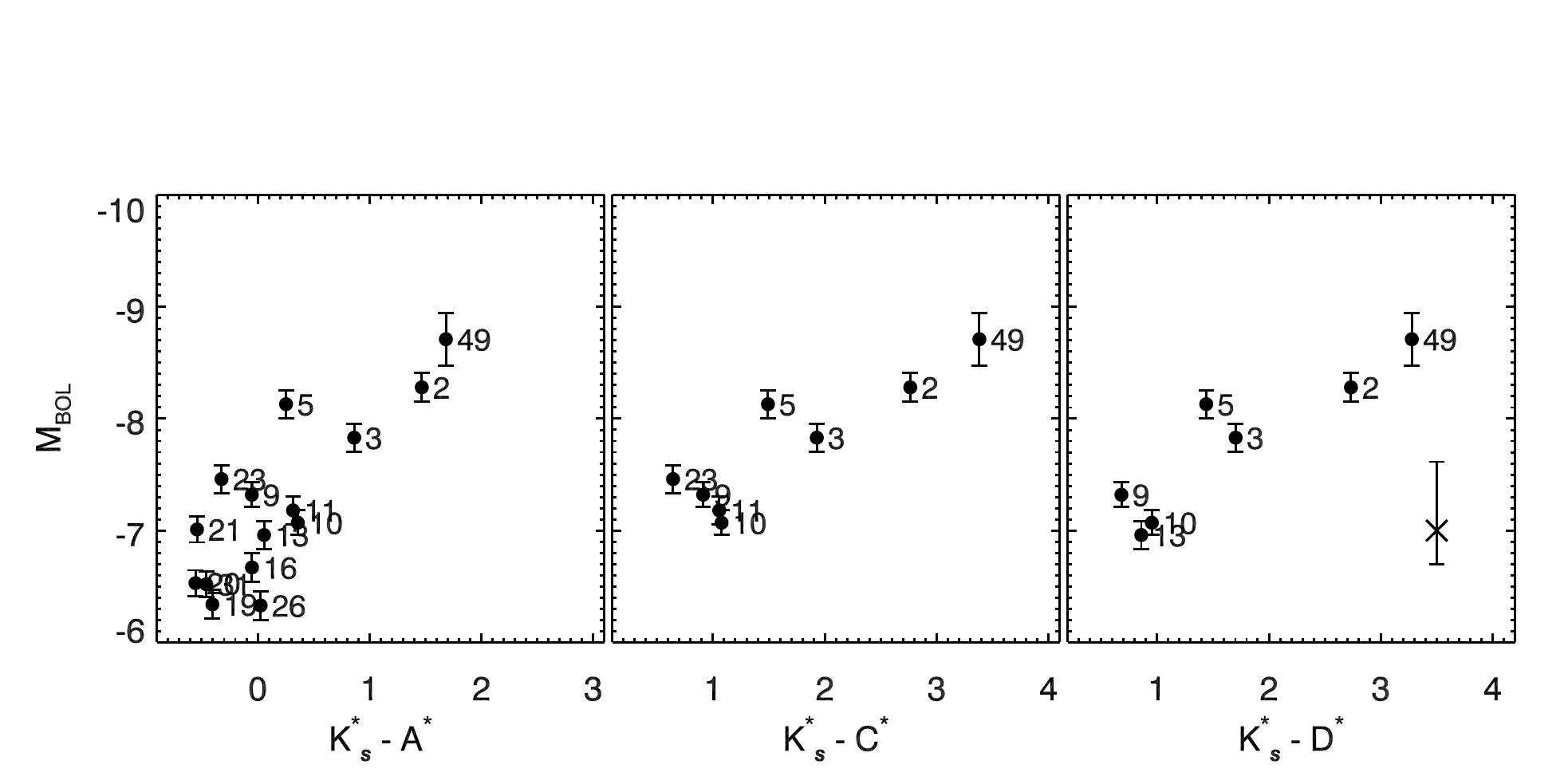}
  \caption{{\it 2MASS-MSX} colour-magnitude diagrams. The plots show
  the cluster stars' bolometric absolute magnitude against {\it
  dereddened} $K_{S}$-($A,C,D$) colours. Errors in $M_{\rm Bol}$ are
  shown which discount the error in cluster distance -- when included
  the average error size is indicated by the error bar in the
  bottom-right of the far-right panel. Errors in colour are of order
  the size of the plotting symbols.  }
  \label{fig:m-kc}
\end{figure}
\begin{figure}[p]
  \centering
  \includegraphics[width=10cm,bb=10 0 646 556]{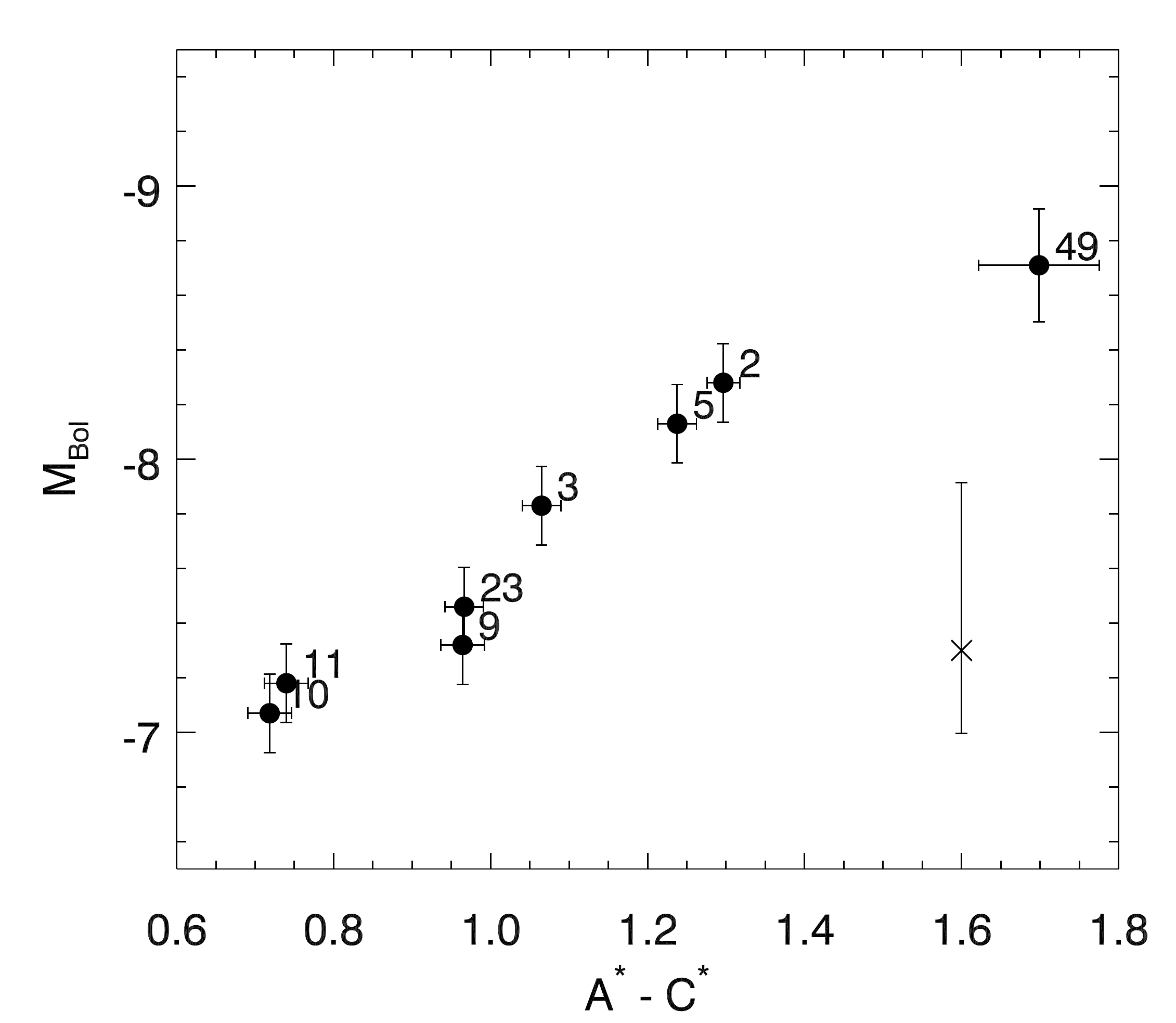}
  \caption{Absolute magnitude versus unreddened $A-C$ colour,
  illustrating the trend of increasing [8]-[12] excess with luminosity
  for the cluster RSGs. The errors in $M_{\rm Bol}$ do not include the
  uncertainty in the cluster distance, which is illustrated by the
  error-bar in the lower right. }
  \label{fig:m-ac}
\end{figure}

In the mid-IR photometry of {\it MSX}, many stars show evidence of
significant excess emission, particularly at 12\microns. This mid-IR
excess is common to RSGs, and is due to the large amounts of
circumstellar oxygen-rich dust produced in this high mass-losing
phase. The excess emission is illustrated in the dereddened {\it
2MASS-MSX} colour-magnitude diagrams shown in Fig.\ \ref{fig:m-kc}. As
these unreddened photospheric colours should be approximately zero,
positive {\it 2MASS-MSX} indices are indicative of dusty circumstellar
ejecta.

\citet{Whitelock94} showed that the $K - [12]$ colour from {\it IRAS}
photometry was directly proportional to mass-loss rate for AGB
stars. This colour may be the most effective diagnostic for RSG
mass-loss rates, as it will be influenced by the broad silicate
emission feature at $\sim$9--12\microns\ often seen in RSG spectra
\citep{Buchanan06}. As RSGs' mass-loss rates are roughly proportional
to their luminosity \citep{v-L05}, in Fig.\ \ref{fig:m-kc} we plot the
bolometric magnitude of the stars against their {\it K-MSX}
colours. Each plot shows the expected $\sim$linear trend of redder
colours for increasingly brighter stars, which was also seen in
a sample of Galactic RSGs by \citet{Massey05} when using the data
from \citet{Josselin00} and re-deriving the stellar distances.

Another effective measure of mass-loss rate may be the [8]-[12] ({\it
MSX} $A$-$C$) colour. This measures not just the mid-IR excess, but
specifically the amount of excess caused by the broad silicate dust
feature. Figure \ref{fig:m-ac} plots absolute bolometric magnitude
against this colour, and shows a clear relation of increasing
12\micron\ excess with increasing luminosity. A comprehensive study of
the empirically-derived mass-loss rates of both the RSG clusters is
beyond the scope of this work, and will be the subject of a future
paper.



\section{Discussion} \label{sec:disc}
\subsection{Initial mass}
To determine the initial mass of the cluster, we use two independent
methods. In Sect.\ \ref{sec:virial} we use our high-precision radial
velocity measurements to infer the cluster's virial mass, under the
assumption that it is in dynamical equilibrium. Secondly, in Sect.\
\ref{sec:evol} we compare this measurement with that determined from
simulations of clusters with large numbers of RSGs, using stellar
evolution models.

\subsubsection{Virial mass} \label{sec:virial}
Following the method of \citet{Mengel02}, the mass of the cluster can
be measured under the assumption that it is in dynamical
equilibrium. Below, we use the dispersion in radial velocity from the
high-resolution spectra to estimate the dynamical mass of RSGC2. As a
caveat, we note that recent work by \citet{B-G06} suggests that young
massive clusters ($\la 50$Myr, $\ga 10^{4}$\msun) can be {\it out} of
virial equilibrium. This is caused by the violent relaxation of the
cluster following the ejection of the left-over interstellar natal
material by the first supernovae. As a result, the cluster mass can be
overestimated by factors of up to $\sim$3 if it is incorrectly assumed
to be virialized. With this in mind, we consider our derived virial
mass to be an order-of-magnitude estimate.

From the radial velocity dispersion $\sigma_{v}$, we can
estimate the dynamical cluster-mass $M_{\rm dyn}$ from the relation,

\begin{equation}
M_{\rm dyn} = \frac{ \eta \sigma^{2}_{v} r_{\rm hp} }{G}
\label{equ:mdyn}
\end{equation}

\noindent where $r_{\rm hp}$ is the half-light radius of the cluster,
$G$ is the gravitational constant, and $\eta$ is a constant which
depends on the density and $M/L$ as a function of radius, and is
typically in the range 5-10 \citep[for a review of the parameter
$\eta$ see Introduction of][]{Mengel02}. Below, we discuss our
measurements of the parameters in this relation and estimate the
dynamical mass of RSGC2.

\paragraph{Velocity dispersion:} As mentioned in Sect.\
\ref{sec:anal}, the method we use to determine accurate radial
velocities from the CO bandhead feature was noted by \citet{Figer03}
to introduce a systematic offset as a function of \EW\ when applied to
Red Giants. This offset was determined from measurements of template
Red Giants with well-known radial velocities. From the few template
spectra of RSGs, it appears that the effect of this systematic
uncertainty is less than $\pm$2\kms\ in our data, over the full range
of equivalent widths.

\begin{figure}[h]
  \centering
  \includegraphics[width=10cm,bb=60 10 600 490]{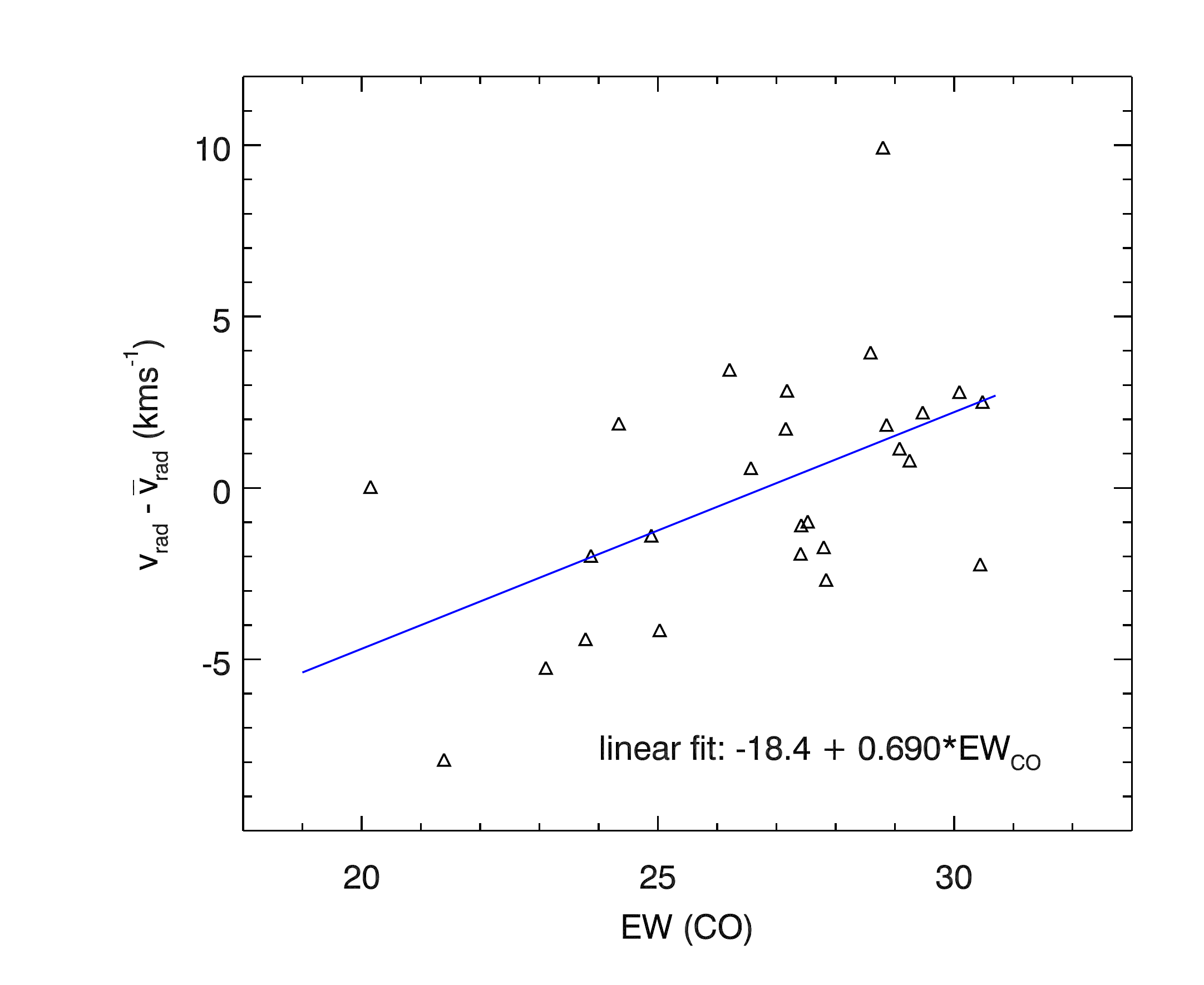}
  \caption{Cluster velocity dispersion as a function of equivalent
  width of the CO bandhead feature. The slight trend may be an
  artifact of the method used to determine the radial velocities of
  the stars. }
  \label{fig:vcorr}
\end{figure}

To investigate this further, in Fig.\ \ref{fig:vcorr} we plot the
radial velocity of the `cluster-members' against their equivalent
widths. The radial velocities have been shifted by the mean velocity
of all the stars, to illustrate the width of the dispersion. There is
an apparent trend of velocity with $EW_{\rm CO}$, which has a Pearson
correlation coefficient of 0.5 when a linear relationship is assumed,
and is plotted over the data. There is no reason to expect a real
trend to exist between velocity and equivalent width, therefore we
suspect that this relation may be an artifact of our
velocity-measuring method.

\begin{figure}[p]
  \centering
  \includegraphics[width=8.5cm,bb=60 10 480 420]{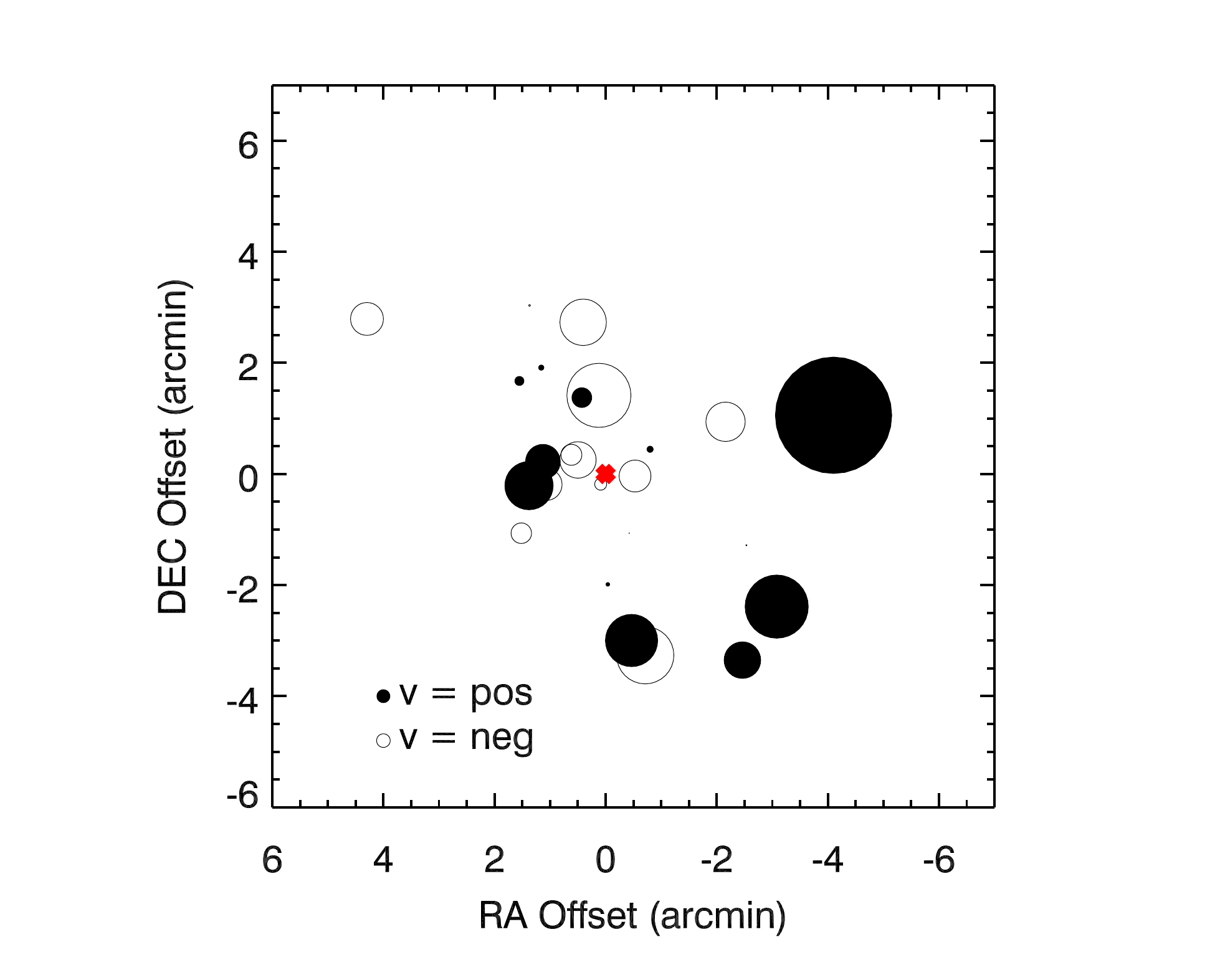}
  \caption{Illustration of the velocity dispersion of the stars in the
  cluster, when the trend with equivalent width has been corrected for
  and the mean cluster velocity subtracted. Symbol sizes are
  proportional to absolute velocity of each star, open symbols
  representing negative velocities and filled symbols positive
  velocities. The cross denotes the cluster `centre', defined as the
  mean of the positions of the stars. }
  \label{fig:vdisp}
\end{figure}
\begin{figure}[p]
  \centering
  \includegraphics[width=8.5cm,bb=60 10 550 420]{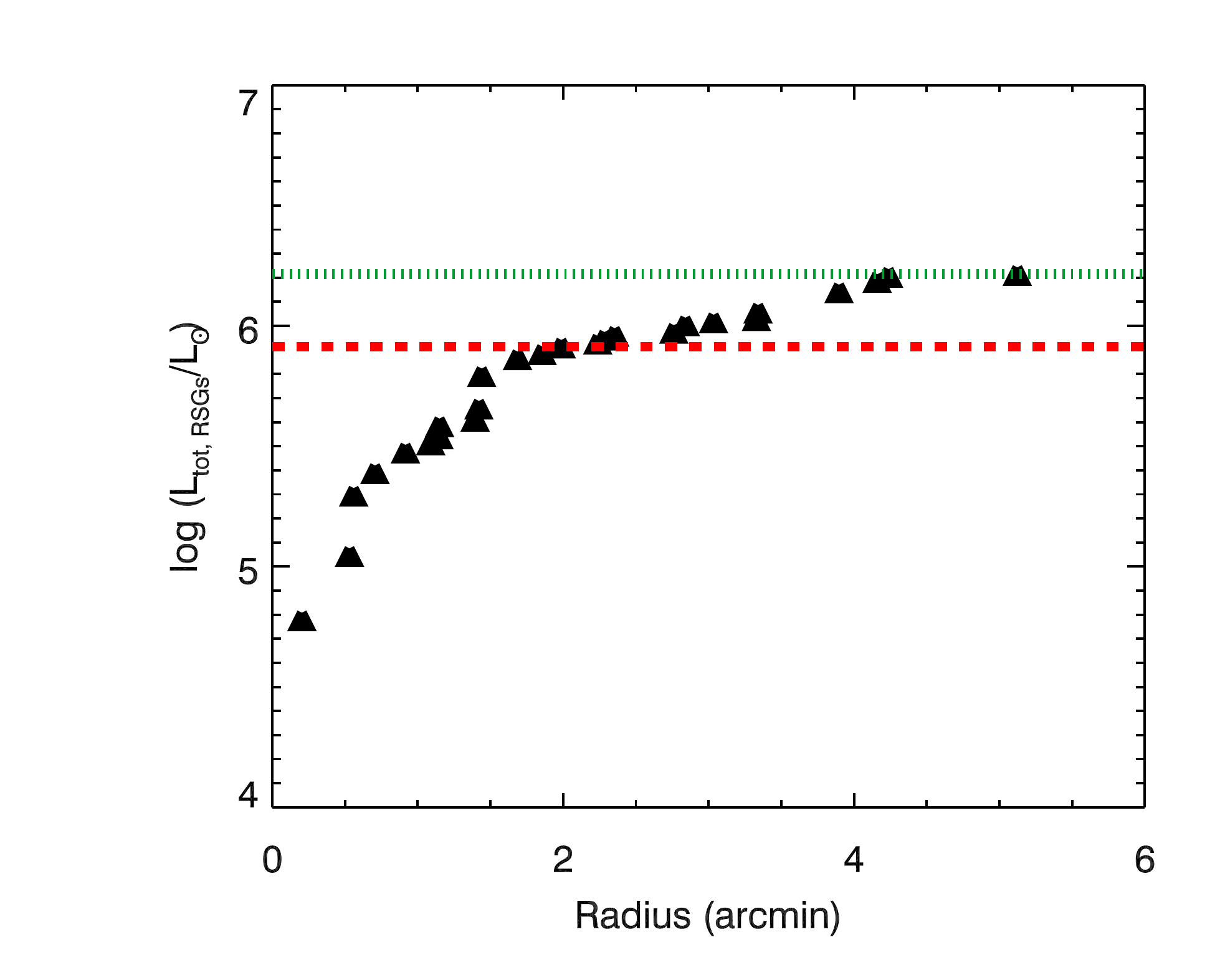}
  \caption{Cumulative luminosity of the cluster, assuming the
  mean of the star positions to be the cluster centre. The maximum
  light and half-light are marked with dotted and dashed lines
  respectively. }
  \label{fig:halflight}
\end{figure}

We investigate the effects of taking two differing estimates for the
velocity dispersion: the {\it maximal} (observed) dispersion, and the
{\it minimal} dispersion when the linear relation shown in Fig.\
\ref{fig:vcorr} is corrected for. In each of these estimates we have
subtracted the internal uncertainty in each measurement ($\pm$1\kms)
in quadrature. We measure the uncorrected data, i.e.\ that plotted in
Fig.\ \ref{fig:vcorr}, to have a 1$\sigma$ dispersion of 3.4\kms;
while the `corrected' velocities have 1$\sigma$ dispersion of
2.8\kms. Therefore, we estimate that if any systematic uncertainty
exists, its effect is at most $\pm$20\%. We illustrate the `corrected'
velocity dispersion of the cluster in Fig.\ \ref{fig:vdisp}. The plot
seems to be a trend of postitive velocities in the south-west and
positive in the north-east, consistent with a rotating
self-gravitating cluster

\paragraph{Half-light radius:} 
As this cluster is much more `open' than its neighbour RSGC1, the
half-light radius is more difficult to define. To measure this
quatity, we make the assumption that the RSGs are representative of
the density profile of the cluster. Should mass-segregation exist in
the cluster, this will be an underestimate. Figure \ref{fig:halflight}
plots the cumulative luminosity distribution of the stars in the
cluster, assuming that the mean of the star positions (18$^{h}$ 39$'$
17.9$''$, -6\deg 2$'$ 3.3$''$) is the cluster centre. The plot
indicates the cluster half-light radius is around (1.9$\pm$0.3)$'$. It
was found that adjusting the position of the cluster centre by
$\pm$0.5\arcmin\ did not significantly affect this value with respect
to the quoted uncertainty.

\paragraph{Density parameter, $\eta$:} 
As we do not have extensive data on many stars in the cluster (the
RSGs are the only stars of which we can be sure are cluster members),
measuring the $\eta$-parameter is beyond the scope of this work. For
now, we use the canonical value of 10, which was shown be
\citet{Spitzer87} to be valid for a range of models.

Using these values and their associated uncertainties, we find a
dynamical cluster mass for RSGC2 of (6$\pm$4)$\times10^{4} (\eta/10)
$\msun, assuming the kinematic cluster distance of 5.83kpc. Below, we
will compare this value to that found from comparison with theoretical
isochrones.

\subsubsection{Evolutionary models} \label{sec:evol}
Following the Monte-Carlo method of FMR06, we generate a coeval
cluster of stars with a predefined total cluster mass, and with
initial stellar masses drawn randomly from a sample consistent with a
Salpeter initial mass-function (IMF) \citep{Salpeter55}. For a given
cluster age, we use the synthetic isochrones created from Geneva
non-rotating stellar evolutionary tracks and determine the present-day
luminosities and temperatures of the stars in the cluster. We isolate
the supergiants as those stars with $\log ( L_{\star}/L_{\odot}) >
4.0$, and define the red, yellow and blue supergiants as those with
temperatures ($T_{\rm eff} < 4500$K), ($4500{\rm K}< T_{\rm eff} <
9000$K), and ($T_{\rm eff} > 9000$)K respectively. We then count the
numbers of RSGs for each simulated cluster for a given age and initial
mass. Each simulation is repeated $10^{3}$ times to reduce statistical
error. We note that while the statistical uncertainty in the mean
number of RSGs per model cluster is negligible, the 1$\sigma$ standard
deviation of the mean is around 20\% (FMR06). Hence cluster parameters
which result in $N_{\rm RSG}$ = 26 can produce 20$\la N_{\rm RSG}
\la$32 for a given trial.

\begin{figure}[h]
  \centering
  \includegraphics[width=10cm]{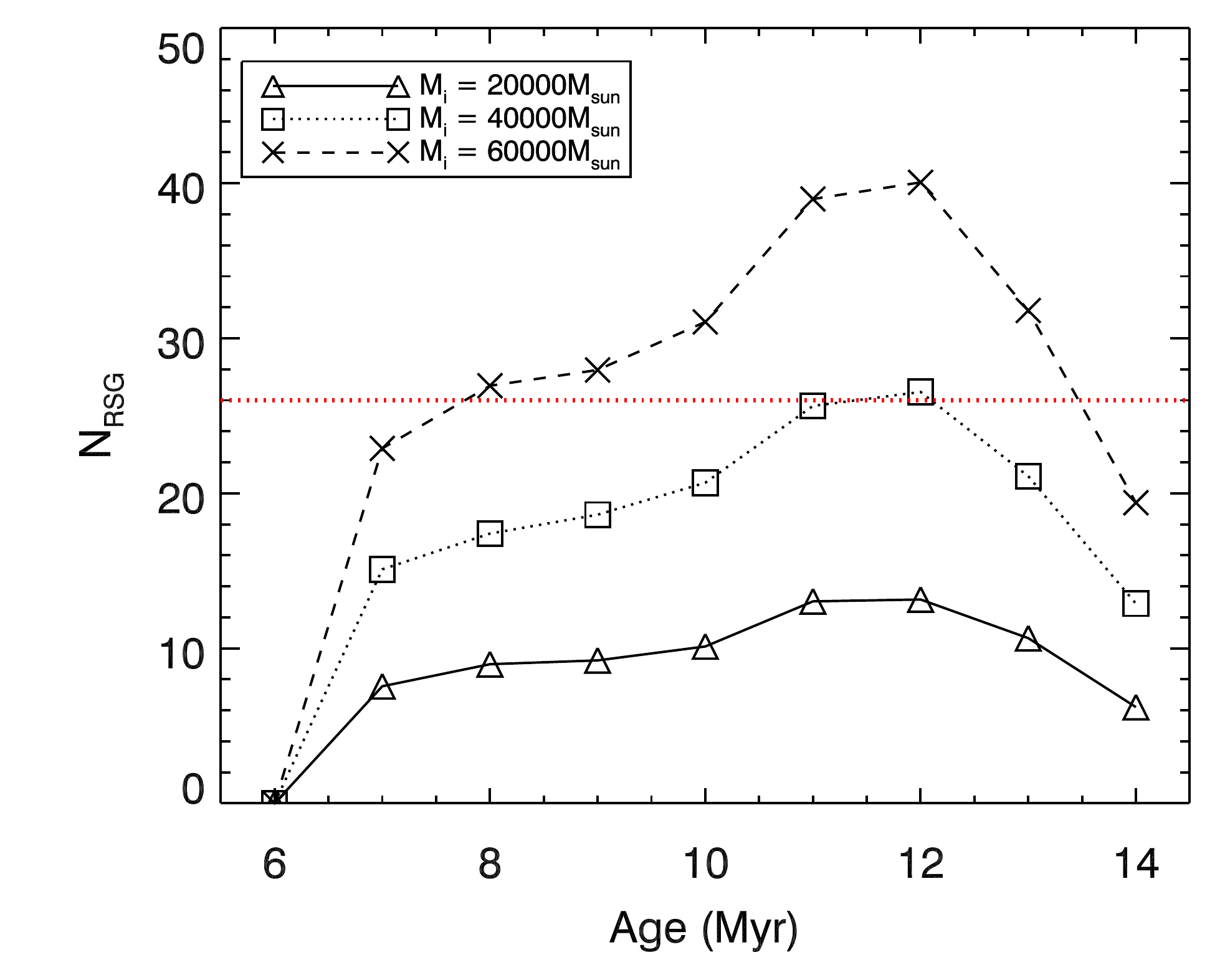}
  \caption{The number of RSGs in a cluster as a function of age for
  three different initial cluster masses, as calculated from the
  non-rotating Geneva models of \citet{Schaerer93}. The observed
  number of 26 RSGs is indicated by the red dotted line.}
  \label{fig:nrsg_mass}
\end{figure}

In Fig.\ \ref{fig:nrsg_mass} we plot the number of RSGs in a coeval
cluster of a given initial mass as a function of age. For this plot we
use the isochrones with solar metallicity and canonical mass-loss
rates of \citet{Schaerer93}. At ages below $\sim$7Myr, very few RSGs
are present. The massive stars which have evolved off the MS
experience high mass-loss in the BSG phase which prevents their
evolution to the red. RSGs begin to appear when those stars with
$M_{initial} \sim 25$\msun\ finish core-hydrogen burning. The number
of RSGs then falls off rapidly above $\sim$14Myr as the stars massive
enough to become RSGs exhaust their nuclear fuel. Hence, the
likelyhood of observing a cluster of 26 RSGs is much higher for
cluster ages in the range 7-13Myr.

\begin{figure}[p]
  \centering
  \includegraphics[width=10cm]{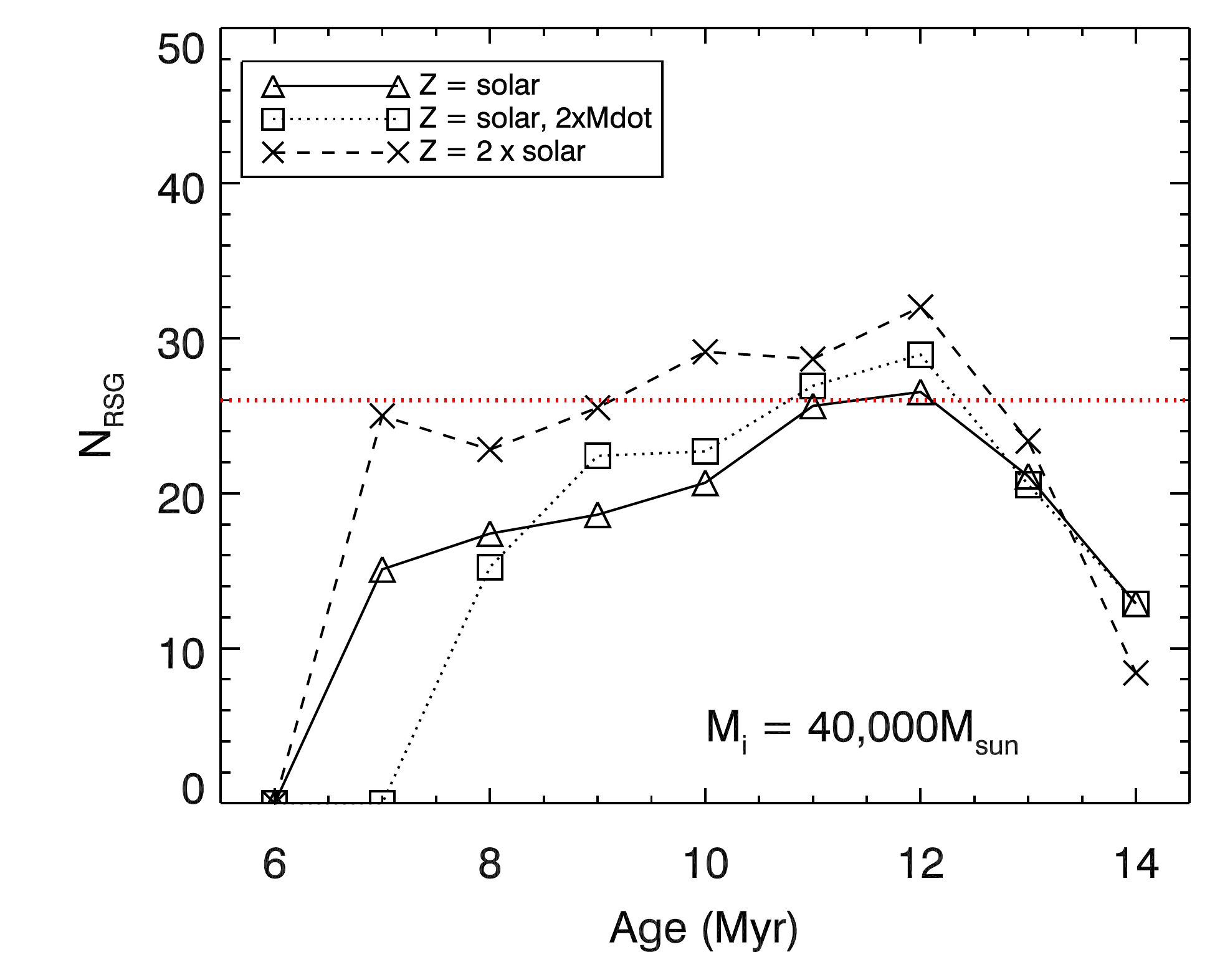}
  \caption{The number of RSGs in a cluster as a function of age for an
  initial cluster mass of $M_{initial} = 4 \times 10^{4}$\msun, as
  calculated from the non-rotating models of \citet{Schaerer93} and
  \citet{Meynet94}. Three different evolutionary tracks are
  investigated: solar metallicity with canonical mass-loss rates;
  solar metallicity with doubled mass-loss rates; and twice-solar
  metallicity. The observed number of 26 RSGs is indicated by the red
  dotted line.}
  \label{fig:nrsg_model}
\end{figure}

\begin{figure}[p]
  \centering
  \includegraphics[width=10cm]{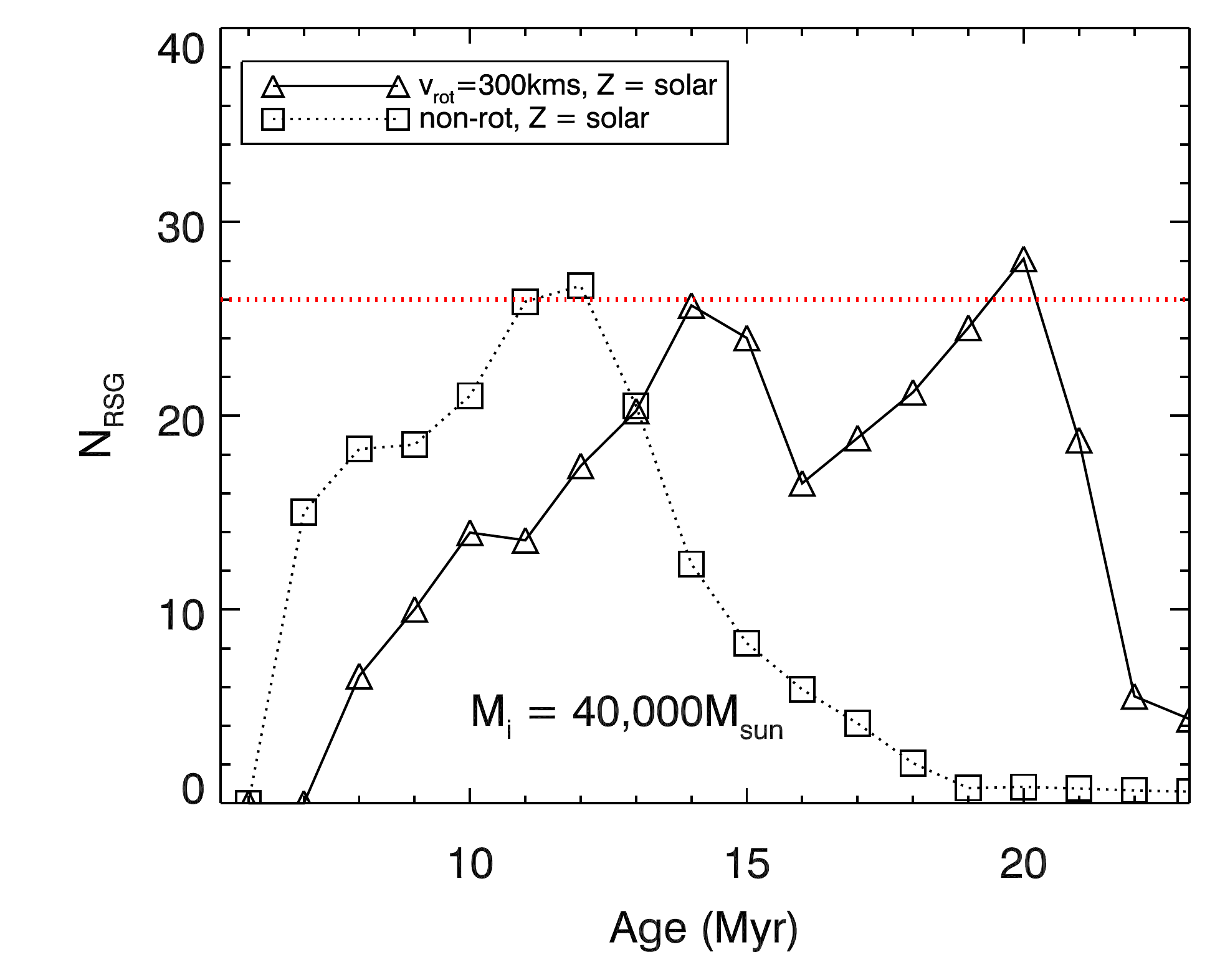}
  \caption{Same as Fig.~\ref{fig:nrsg_model}, comparing the
  non-rotating -- solar-metallicity models of \citet{Schaerer93} with
  the fast-rotating, solar-metallicity models of \citet{Mey-Mae00}.}
  \label{fig:nrsg_model_rot}
\end{figure}

Figure \ref{fig:nrsg_mass} illustrates that for coeval clusters, only
those with $M_{initial} \sim 4 \times 10^{4}$\msun\ can produce
numbers of RSGs in excess of 26, the number we observe in this
cluster, for the evolutionary models used in Fig.\
\ref{fig:nrsg_mass}. In Fig.\ \ref{fig:nrsg_model} we investigate the
effect of using isochrones generated with different evolutionary
tracks. We plot the results of using isochrones with solar metallicity
and canonical mass-loss rates (as used in Fig.\ \ref{fig:nrsg_mass}),
solar metallicity with doubled mass-loss rates \citep{Meynet94}, and
twice-solar metallicity. Whilst the different models produce slightly
different results, in the likely range of cluster ages of 7-13Myr the
model differences are smaller than the statistical variations of
individual simulations. We therefore consider this to be a negligible
source of uncertainty in this estimate of the cluster mass.

Much progress has been made in recent years in incorporating rotation
in stellar structure codes \citep[see review of][]{M-M00}, so a
discussion of the effects of including rotation on our analysis seems
warranted. \citet{H-L00}, \citet{Mey-Mae00}, and \citet{Heger00}
studied specifically the effect of rotation on stars in the initial
mass range relevant to this work, i.e.\ $M_{\star} \sim$
15-25\msun. The broad result was that rotationally-enhanced mixing
increases the chemical homogeneity of the star, leading to larger
helium cores, higher luminosities ($\Delta \sim$0.25dex) and lower
effective temperatures ($\Delta \sim$400K) of RSGs. In addition, stars
spent longer on the main-sequence ($\sim$12\%), due to the decreased
effective gravity causing the star behave as a non-rotating star with
lower initial mass.

In Fig.~\ref{fig:nrsg_model_rot} we investigate the effect of using
the contemporary Geneva models which include stellar rotation. The
rotational-velocity grids of these models are, as yet, not
complete. Here we use the $v_{i}$=300\kms\ models, which are likely
too large for this mass-range and metallicity. However they serve to
investigate the impact of stellar rotation on our analysis. As the
impact of rotation is greatest on the evolution of massive stars, due
to its effect on the mass-loss behaviour, Geneva models are only
computed down to 9\msun. To construct isochrones we spline together
the massive, rotating models with the non-rotating models of mass
$<$9\msun.

The Figure shows that, in the early part of the diagram ($\la$15~Myr),
the rotating models lag the non-rotating models, due to the longer
time spent on the main-sequence. At later times, the rotating models
continue to produce RSGs long after the non-rotating models. This can
be understood as a combination of longer lifetimes, and the stars'
increased luminosity and decreased $T_{\rm eff}$, enabling them to
spend longer in the RSG `zone', as defined by our somewhat arbitrary
thresholds of $\log (L/$\lsun) $\ge$4.0 and $T_{\rm eff}
\le$4500~K. The inclusion of rotation does not affect the inferred
lower limit to the initial cluster mass of $M_{initial} \sim 4 \times
10^{4}$\msun. 

From isochrone fitting (see next section), we are able to constrain
the age of the cluster to $12\pm1$~Myr (non-rotating models) and
$17\pm3$~Myr (fast-rotating models). From these results, we estimate
an `evolutionary' cluster mass for RSGC2 of $M_{\rm ev} = (4 \pm 1)
\times10^{4}$\msun. The uncertainty takes into account the statistical
variations of the Monte-Carlo method, and the error in the cluster
age. The estimate compares well to the dynamical mass of $M_{\rm dyn}
=$ (6$\pm$4)$\times10^{4} (\eta/10) $\msun.

\subsubsection{The effect of cluster non-coevality}
The mass derived above assumes that the stars were created in a coeval
starburst. The large extent of the association ($\sim$10pc at a
distance of 5.83kpc), as well as the large luminosity spread (see
Sect.\ \ref{sec:age}), may suggest a sustained starburst phase of
several million years. While the large size may be explained by
expansion due to non-virial equilibrium \citep[][see Sect.\
\ref{sec:virial}]{B-G06}, and the luminosity spread due to
short-comings in evolutionary models \citep[][Sect.\
\ref{sec:age}]{M-O03}, we nonetheless discuss the effect of non-coeval
star-formation on our derived total mass.

The effect of cluster non-coevality would be to convolve the curves
shown in Fig.\ \ref{fig:nrsg_mass} with a smoothing function
characterized by the length of the starburst phase. Thus, as long as
the starburst occured on timescales much shorter than the mean age of
the cluster, it would not significantly affect the number of RSGs
observed at any one time. For an extended starburst phase of order the
inferred age of the cluster, the number of RSGs at any one time for a
given cluster mass would decrease. Hence, a prolonged starburst would
imply a larger cluster mass than derived here. The presence of
main-sequence O stars or WRs, which have lifetimes of $\sim$3Myr, or
low-mass AGB stars with lifetimes of $\gg$20Myr, would imply a
sustained star-forming episode. Observations of such stars would
require precise radial velocity measurements, such as those presented
here, to confirm that they were part of the physical association.

\subsection{Cluster age} \label{sec:age}
\begin{figure}[h]
  \centering
  \includegraphics[width=10cm]{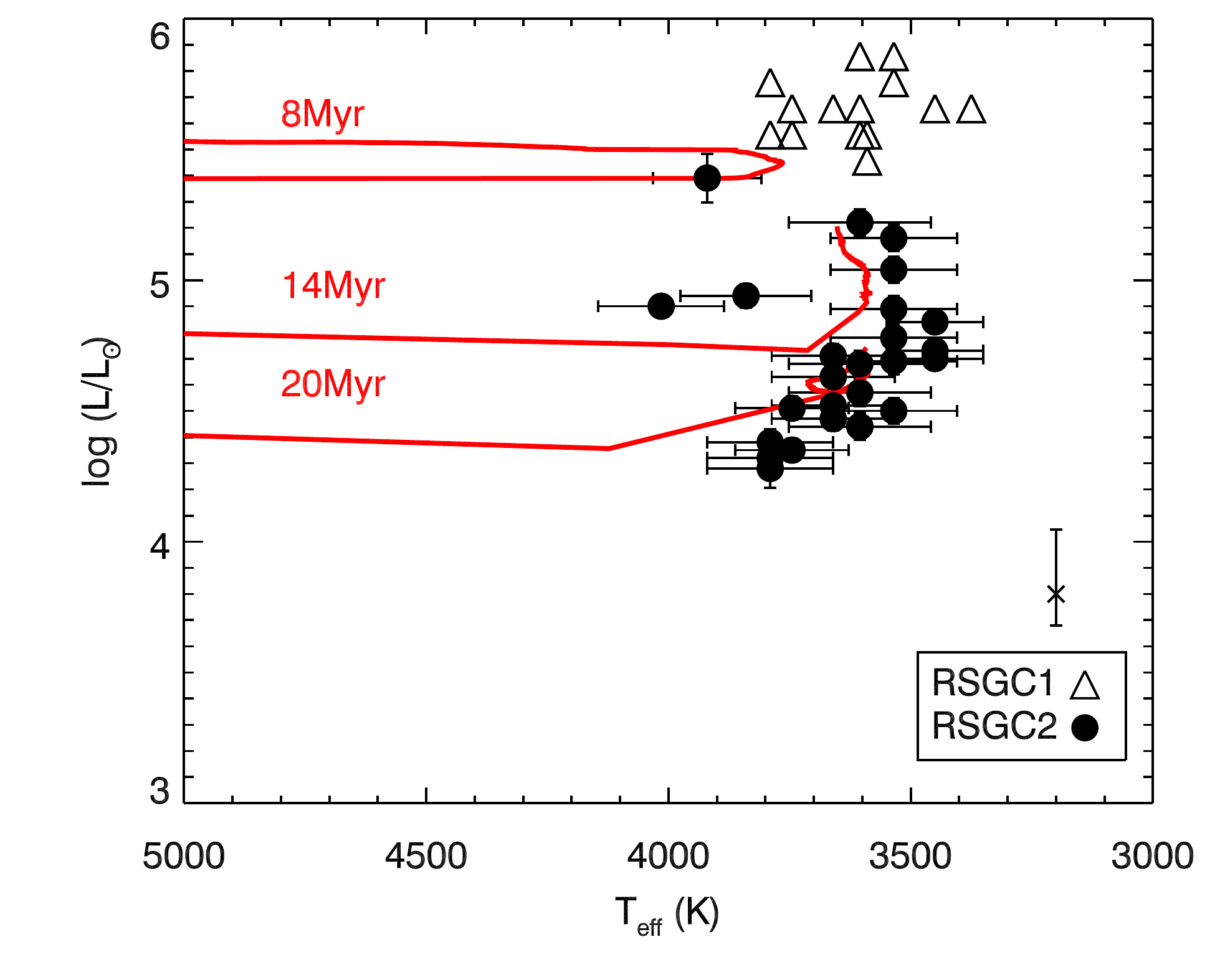}
  \caption{H-R diagram, showing the positions of the stars in the two
  Scutum RSG clusters. Also shown are isochrone fits for each cluster,
  based on rotating Geneva models with solar metallicity, canonical
  mass-loss rates, and initial rotational velocity of 300\kms. As the
  RSGC2 stars are all at the same distance, the uncertainties shown in
  $L_{\rm Bol}$ for these stars do not include the error in the
  cluster distance. The magnitude of the distance error is indicated
  seperately in the bottom-right of the panel.}
  \label{fig:age}
\end{figure}

A novel method for estimating the age of a cluster from the RSG
population was presented in FMR06. They showed that, using the
non-rotating Geneva models, the luminosity range of RSGs for a coeval
cluster changes with age (see their Fig.\ 19). For young ($\la$7Myr)
clusters, the RSGs result from stars which have evolved horizontally
across the HR diagram, meaning that the RSGs occupy a very narrow
luminosity range. For older clusters however, the RSGs -- which result
from stars of lower initial mass -- have a larger luminosity spread,
due to the upturn at the end of the evolutionary path (see Fig.\
\ref{fig:age}).

The RSGC2 stars have luminosities ranging from $\log$($L$/\lsun) = 4.2
$\rightarrow$ 5.2 (see Table \ref{tab:clresults}). This compares to
the larger luminosities and narrower spread of RSGC1, which has
$\log$($L$/\lsun) = 5.0 $\rightarrow$ 5.6. This can be seen clearly in
Fig.\ \ref{fig:age}, which shows the locations of the stars in the two
clusters on a H-R diagram. The luminosities of the stars in RSGC1 are
taken from FMR06, and have been corrected for the slightly greater
distance determined from SiO maser emission by \citet{N-D06} and the
contemporary Galactic rotation curve (see above). Isochrone fits to
the data, again using the fast-rotating Geneva models of
\citet{Mey-Mae00}, illustrate the clear age-difference of the
clusters. An age of $17\pm3$~Myr is consistent for RSGC2, compared to
$8\pm1$~Myr for RSGC1. When non-rotating isochrones are used, the
inferred age of RSGC2 becomes $12\pm1$~Myr, while the age of RSGC1 is
unchanged. The figure shows that, while there may be some overlap in
the absolute uncertainties, there is a clear age-gap between the two
clusters of several Myr.

While it is remarkable how the RSGC2 stars tightly follow the
`hockey-stick' end to the isochrone, no single isochrone reproduces
the luminosity spread of the stars, with the 20Myr rotating isochrone
not extending to the greatest luminosities observed in the
cluster. Taking Fig.\ \ref{fig:age} at face-value, this could imply
that the cluster is non-coeval, and formed over a period of
6Myr. However, it was a well-known problem that the non-rotating
evolutionary models did not reproduce the highest observed
luminosities of RSGs \citep[][]{M-O03}; and while the inclusion of
rotation in evolutionary codes does in general make RSGs redder and
brighter, it is not clear that the difference between observation and
expectation has been completely reconciled. Indeed, the RSGCs may be
the ideal laboratory in which to test these models.

\begin{figure}[p]
  \centering
  \includegraphics[width=16cm,bb=30 00 450 413]{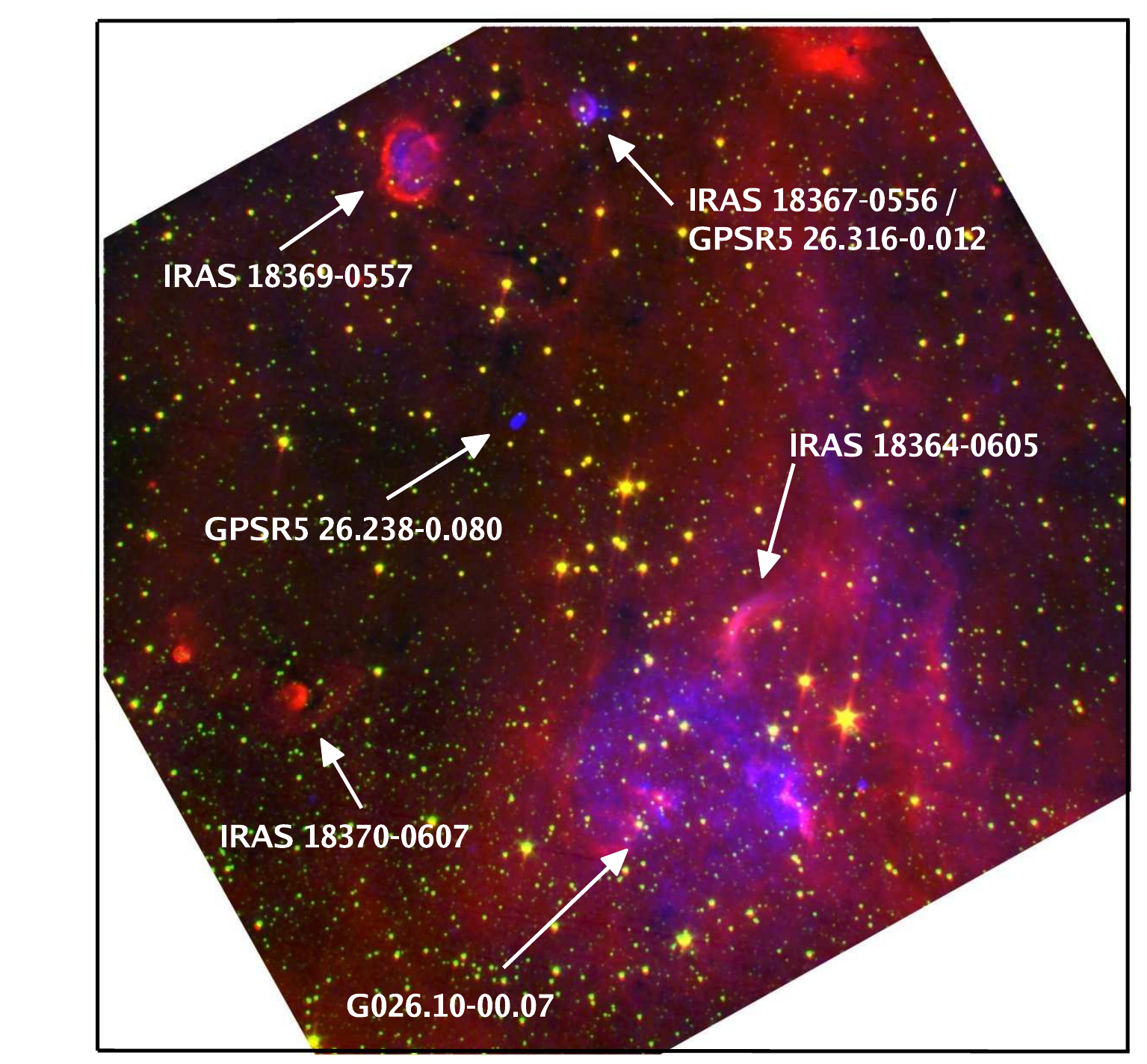}
  \caption{ RGB-composite image of the region around RSGC2. The image
  shows the {\it Spitzer/GLIMPSE} bands 8.0\microns\ (red),
  4.5\microns (green), and {\it MAGPIS}-20cm (blue). Sources
  identified in radio surveys, as well as the likely origin of several
  {\it IRAS} point-sources, are marked on the image. }
  \label{fig:comp}
\end{figure}

\subsection{Unidentified objects near RSGC2: evidence of recent
  starburst activity at the base of the Scutum-Crux arm. }
  \label{sec:others}

From the derived distances to the two RSG clusters, their separation
is of order 100pc. The proximity of these two remarkable objects to
one another, combined with their similar ages, is perhaps indicative
of a wider-scale starburst episode in the region of the Scutum-Crux
spiral arm. As noted by \citet{N-D06}, the inferred distances for the
objects put them close to where the spiral arms meet the Galactic
bulge, roughly the co-rotation radius of the bulge bar
\citep{Bissantz03}, and in the middle of the proposed high-density
`stellar ring' \citep{Bertelli95}. The physical conditions and gas
dynamics in this region of the Galaxy may precipitate star formation
activity, while the location of the clusters within one of the
co-rotation Lagrangian points \citep{E-G99} may harbour the clusters
from tidal disruption.

As we are looking tangentially along the Galactic arm at the point
where it meets the bulge, it is reasonable to assume that there may be
other evidence of recent star-formation along our line-of-sight
towards the two clusters. Indeed, separate from the cluster stars,
there appear to be further RSGs in the direction of RSGC2. These
objects have CO equivalent widths too large to be giants, but their
radial velocities are inconsistent with being part of the cluster
itself. These stars may be part of smaller clusters formed in a
region-wide starburst phase around $\sim$10--20Myrs ago.

FMR06 discussed the possibility that various unidentified
high-energy/radio sources in the region of RSGC1 were due to recent
supernova activity, although the nonthermal radio-sources have since
been shown to be extra-galactic \citep{T-R06}. Here, we make a similar
discussion of the unidentified sources near RSGC2, using the Galactic
plane survey data of {\it GLIMPSE}, {\it MIPSGAL} and {\it MAGPIS}
\citep{Benjamin03,Carey05,Helfand06}.

Figure \ref{fig:comp} shows a composite of {\it IRAC} channels 2
(4.5\microns) and 4 (8.0\microns), and {\it VLA}-20cm, centred on
RSGC2. The image shows in detail for the first time several radio and
{\it IRAS} point-sources, as well as the H{\sc ii}-region
G26.10-0.07. Below, we discuss the nature of each of these objects, as
well as their relation to RSGC2 and the starburst phase in which it was
created.

\begin{figure}[p]
  \centering
  \includegraphics[height=6.8cm,bb=50 77 400 453,clip]{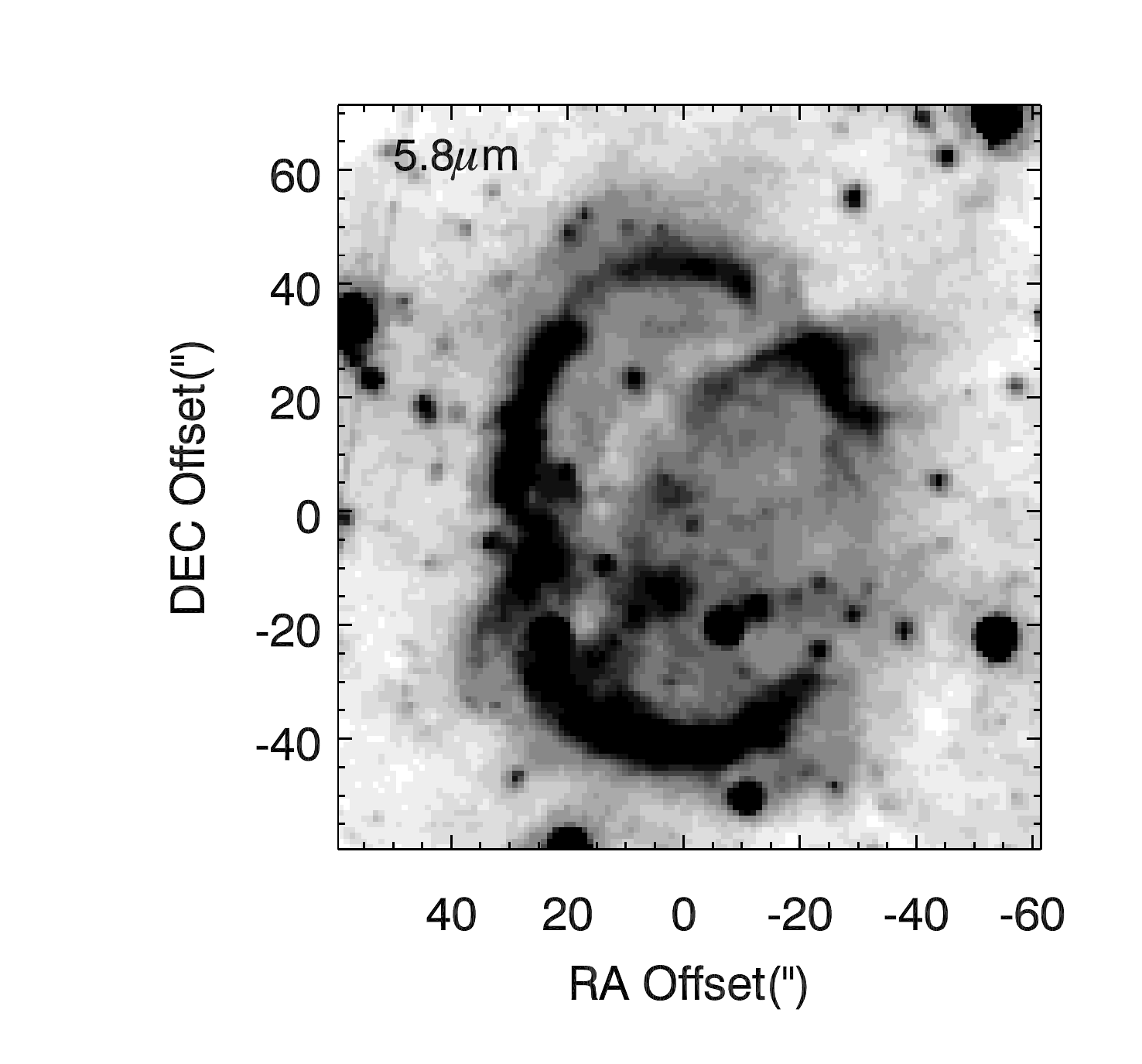}
  \includegraphics[height=6.8cm,bb=123 77 400 453,clip]{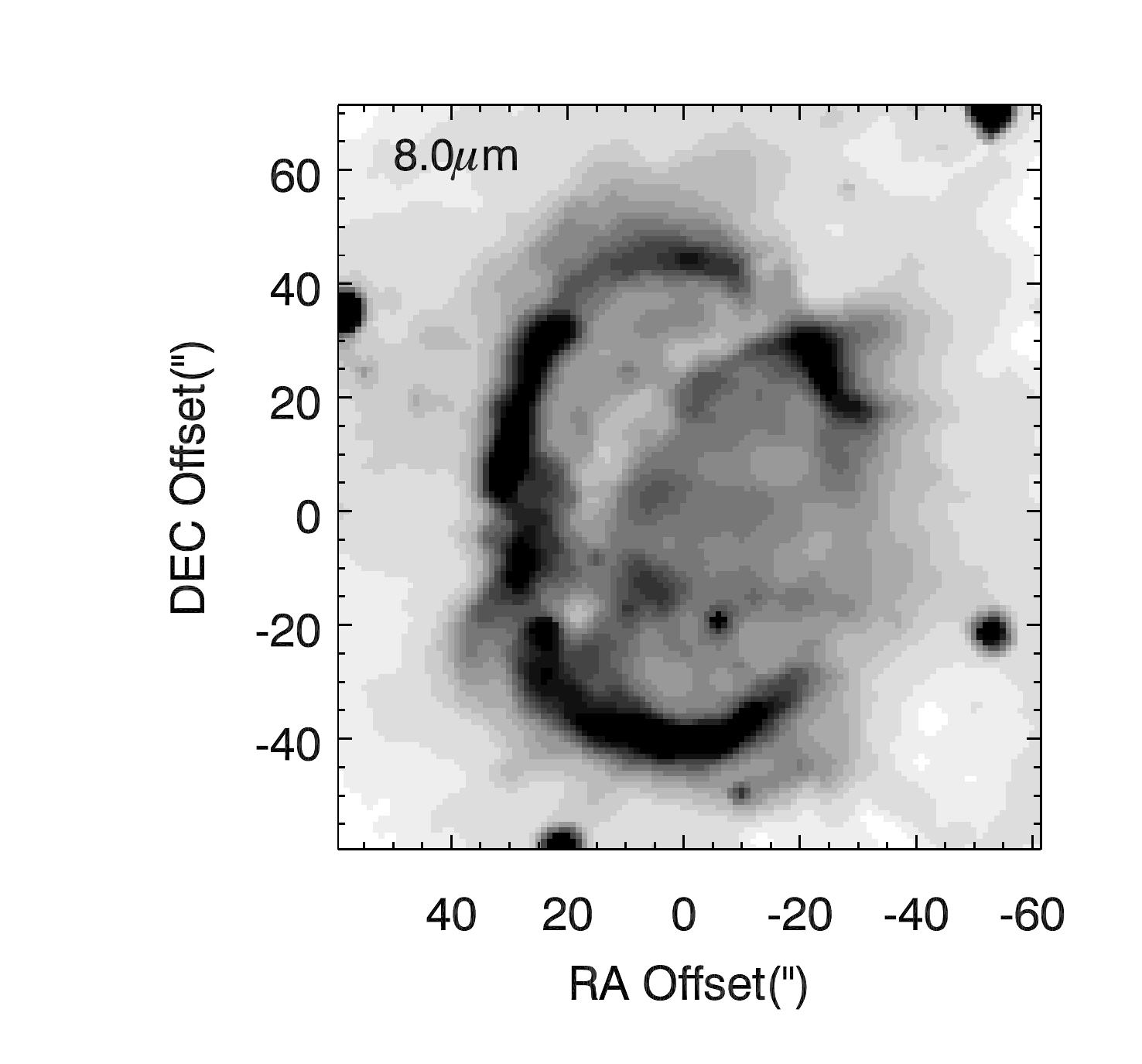} 
  \includegraphics[height=6.8cm,bb=50 -10 400 366,clip]{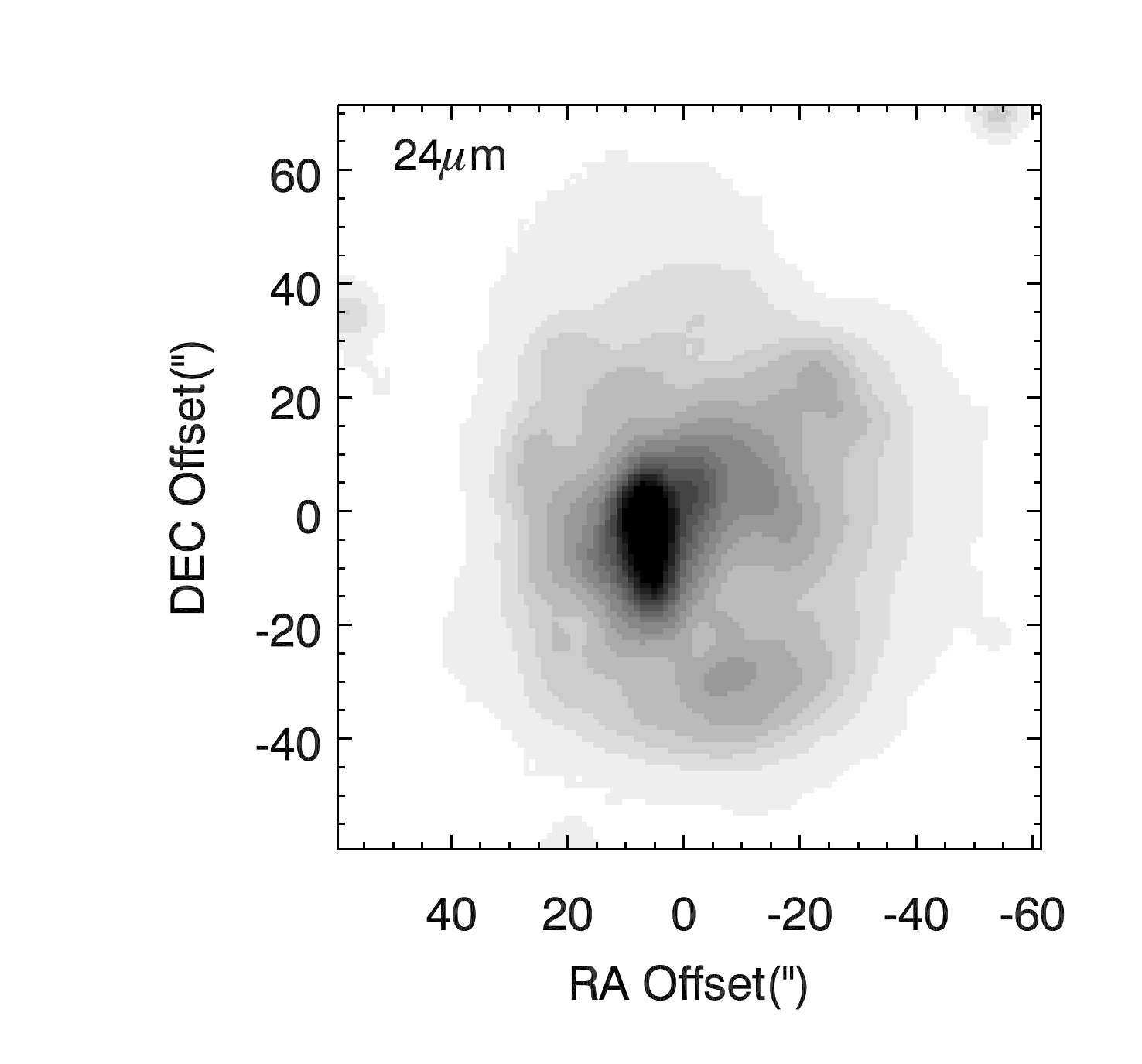}
  \includegraphics[height=6.8cm,bb=123 -10 400 366,clip]{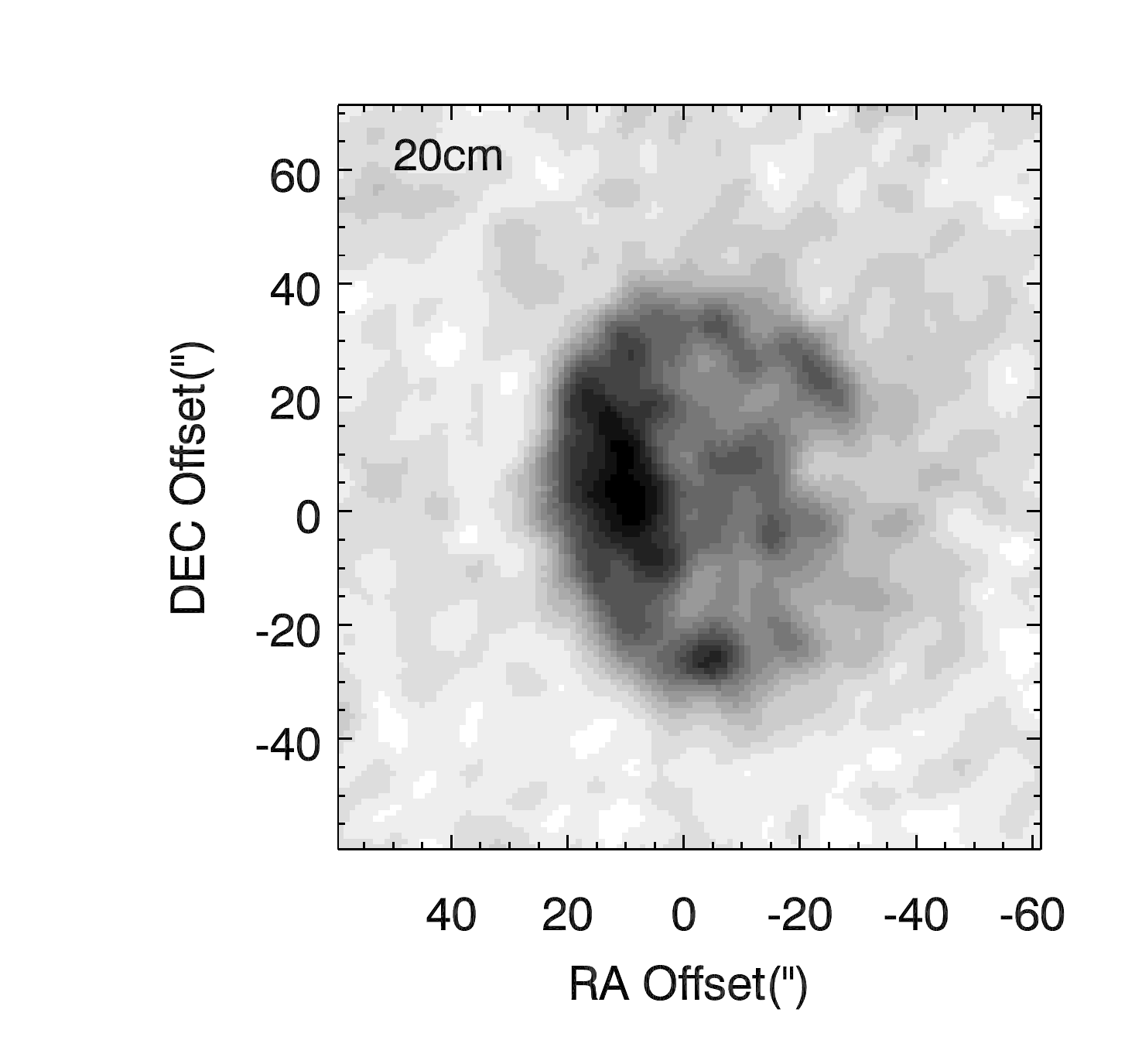}
  \caption{High-resolution images of the IRAS 18369-0557
  source. Top-left: Spitzer/IRAC 5.8\microns; top-right: Spitzer/IRAC
  8.0\microns; bottom-left: MIPS 24\microns; bottom-right: VLA
  20cm. All images are scaled linearly between 0-7$\sigma$ above the
  sky background. }
  \label{fig:snr}
\end{figure}

\paragraph{IRAS 18369-0557}
When seen in detail (see images in Fig.\ \ref{fig:snr}), this object
has the appearance of a discrete ring of material, which seems to peak
in the 5.8 \& 8\microns\ bands. Inside this ring is filled-in with
20cm-emission, and the inner material is also bright at 24\microns\
such that it saturates the {\it MIPSGAL} image. The object is not
detected in {\it 2MASS}, nor is it detected in {\it IRAC} bands 1 or
2. No obvious central point-source is observed in any band. Aside from
the bright ring, there is also a dark arc extending from the SE to the
NW. This arc may be due to cold dust, oriented in some polar outflow
perpendicular to the bright ring. 

That the ring is so bright at 5.8\microns\ but not seen at 4.5\microns,
with the central region peaking at $\sim$24\microns, suggests that the
emission may be due to warm ($\sim$100K) dust, with strong PAH
emission at 5\microns\ in the outer ring. Detailed temperature
modelling of the dust would benefit from mid-IR spectroscopy across
the nebula, such a study is beyond the scope of the current work. 

From the object's appearance and the apparent lack of any central
source, it is tempting to classify the object as a supernova remnant
(SNR). The semi-major axis of the ring is 1.5\arcmin\ across, which at
the distance of RSGC2 corresponds to a diameter of 2.5pc. If we assume
a typical SN expansion speed of $\sim$1000\kms, this would make the
remnant $\sim$2500 years old {\it if} the dust has formed out of the
SN ejecta. It would seem unlikely that 100K dust would survive this
long; by comparison, \citet{Blair07} find that Kepler's SNR, which is
$\sim$4kpc away and $\sim$400 years old, is already very faint at
5.8\microns. A more likely explanation is that the dust was produced
in a pre-SN mass-losing phase of the precursor, and has been heated by
the SN explosion which must have occured less than a few hundred years
ago. Such a situation is seen to be happening in SN~1987A
\citep{Bouchet06}.

It was argued in FMR06, using evolutionary models, that a cluster
similar to RSGC1 should experience SNe explosions every
$\sim$40,000-80,000~yr. For RSGC2, which appears to be 50\% more
massive but a little older, the corresponding timscale is around
50,000yrs. If a SN remnant takes around $10^{4}$~yrs before it becomes
too faint to observe, then it is not unreasonable to assume that we
may observe one recent supernova in a cluster like RSGC2. We note that
at present there is currently no associated high-energy source, and so
classification of the object would benefit from X-ray / $\gamma$-ray
observations. 

\begin{figure}[t]
  \centering
  \includegraphics[height=6.5cm,bb=50 0 400 376,clip]{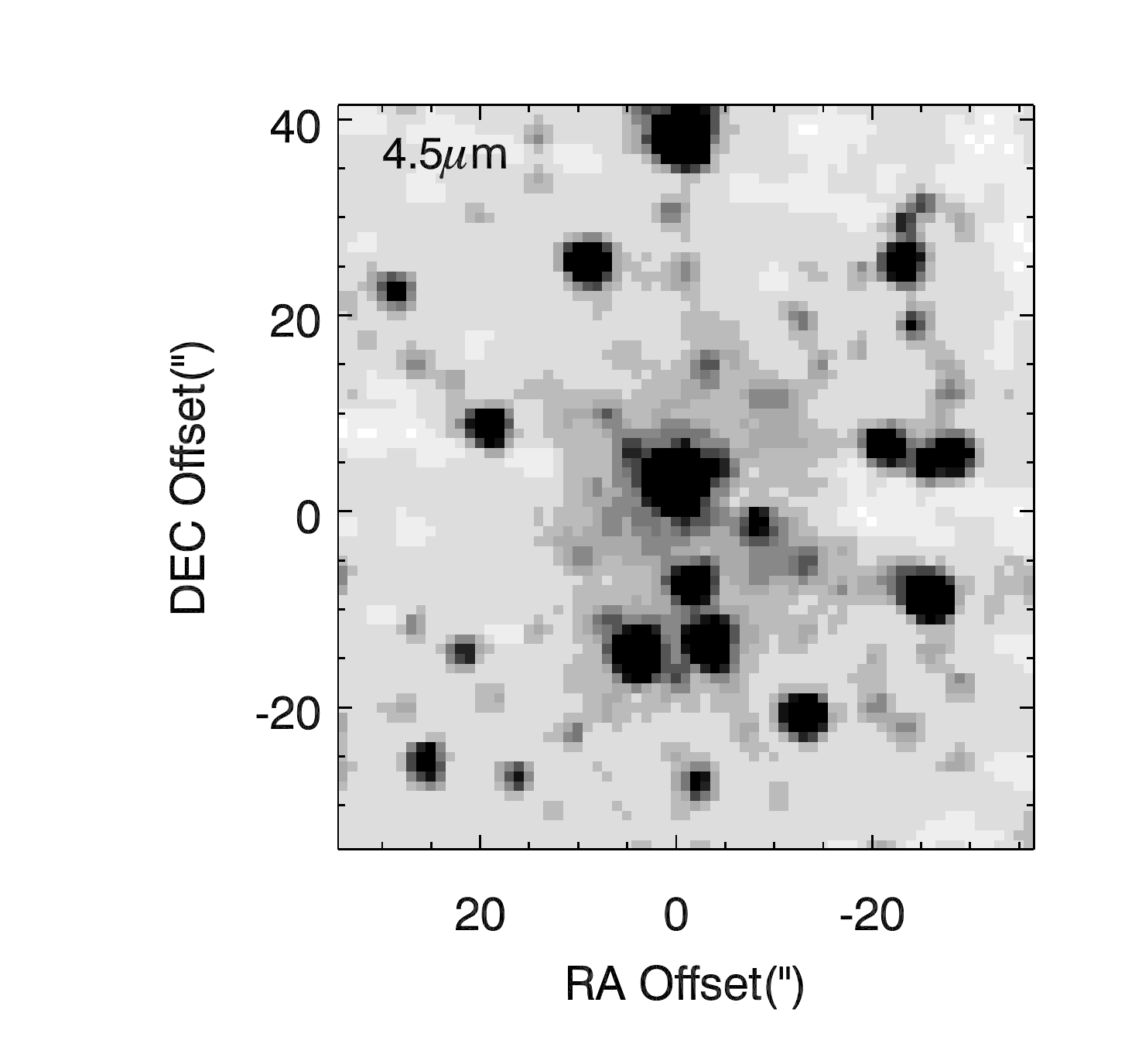}
  \includegraphics[height=6.5cm,bb=123 0 400 376,clip]{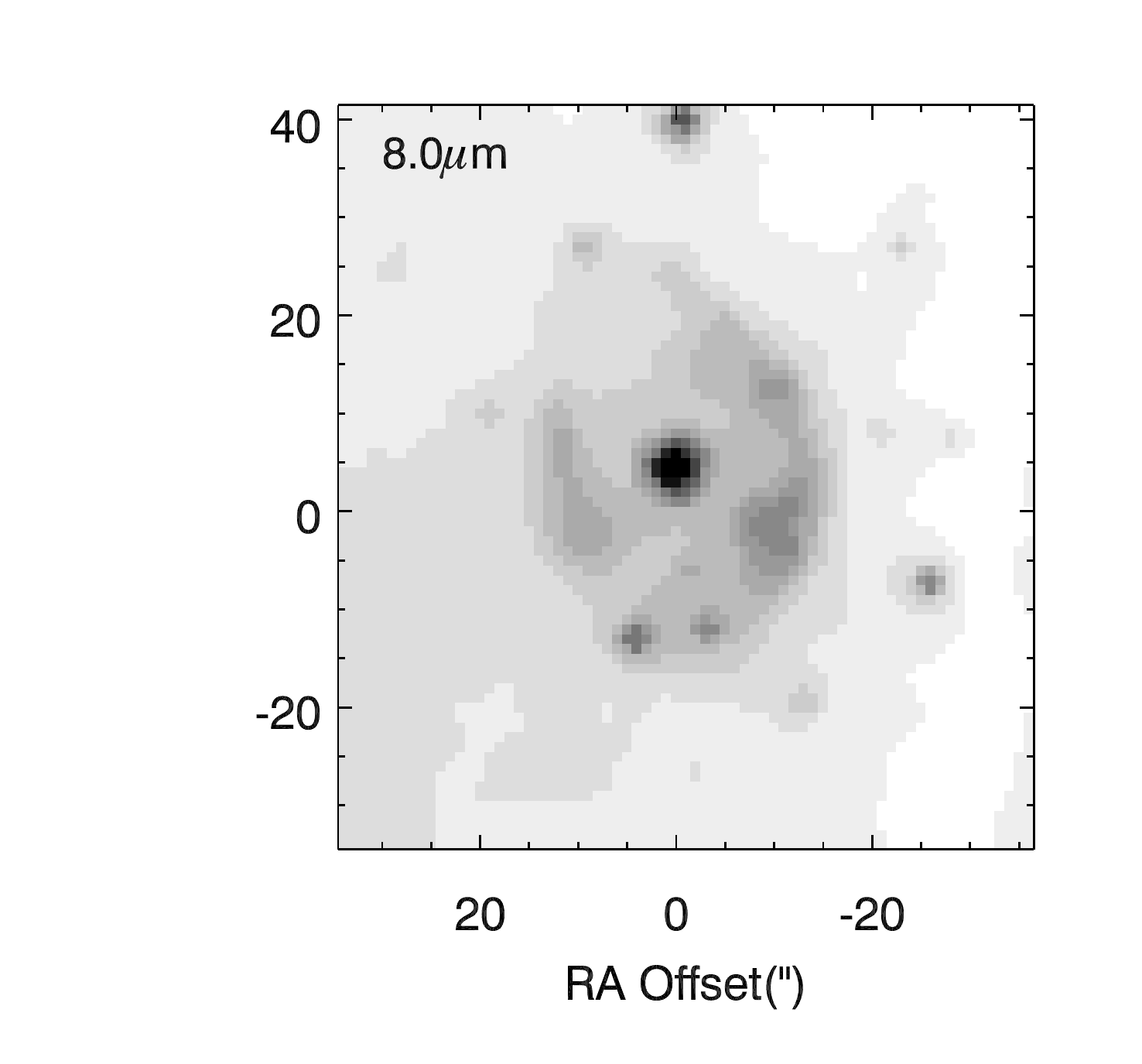}
  \includegraphics[height=6.5cm,bb=123 0 400 376,clip]{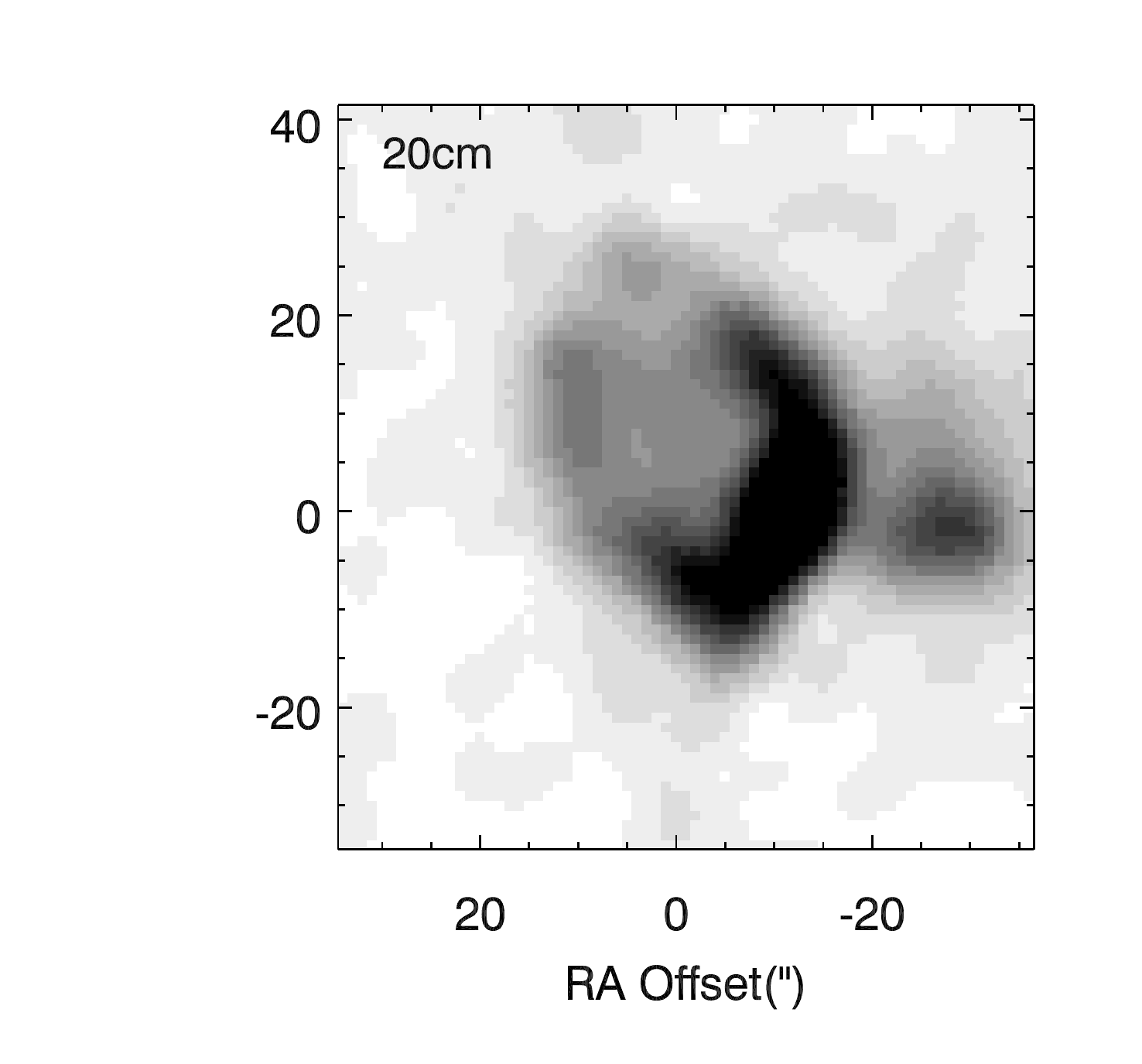}
  \caption{High-resolution images of the IRAS 18367-0556 source. Left:
  Spitzer/IRAC 4.5\microns; centre: Spitzer/IRAC 8.0\microns; right:
  VLA 20cm. All images are scaled linearly between 0-10$\sigma$ above
  the sky background. }
  \label{fig:lbv}
\end{figure}

\paragraph{IRAS 18367-0556 / GPRS5 26.316-0.012}
This object is another radio-bright, 8\microns-bright ring-nebula,
this time with a highly-reddened star at the centre
($H-K_{S}$=1.7). The GLIMPSE and MAGPIS images of the object are shown
in Fig.\ \ref{fig:lbv}. It is so bright at 24\microns\ that it
saturates the corresponding {\it MIPSGAL} image. The object is
reminiscent of an evolved star surrounded by the ejecta of a previous
high mass-losing phase, such as a post-AGB star, an LBV, or a WR
star. Indeed, it is reminiscent of the mid-IR ring-nebulae seen around
candidate LBVs in the {\it MSX} survey \citep{Clark03}, in particular
G26.47+0.02. 

The radio source is detected in the 1.4GHz NRAO VLA Sky Survey (NVSS)
\citep{Condon98}, and has a flux of 65.6$\pm$2~mJy. However, the
source is flagged in the survey as being `complex', possibly due to
the size of the source ($\sim$40\arcsec) being comparable to the size
of the beam (FWHM=45\arcsec). From Fig.\ \ref{fig:lbv} it appears that
the bulk of the radio emission is coincident with the south-western
part of the dust-ring, rather than with the central star. It is
therefore unlikely that the radio emission results from the ionized
stellar wind, as one would expect to see emission from the base of the
wind coincident with the central star. The radio emission could be
explained by a hot central star ionizing the surrounding ejecta, or
the fast wind of a hot phase ploughing into a slower, dusty wind
ejected when the star was cooler.

The object is very reminiscent of the LBV candidate HD~168625
\citep{R-H98}, speculated by \citet{Smith07} to be a Galactic analogue
of the progenitor of SN~1987A based on the recent discovery of an
8\micron\ ring around the star. Clearly, further study of IRAS
18367-0556 is warranted, in particular near-IR spectroscopy to
determine the stellar temperature, compare the star's radial velocity
with the nearby RSG cluster, and to potentially determine abundances
of Fe and $\alpha$-group elements in this region of the Galaxy
\citep[see e.g.\ Introduction of][]{Najarro04}.


\paragraph{GPSR5 26.238-0.080}
This is an extremely compact radio source, with no obvious counterpart
in {\it 2MASS}, {\it GLIMPSE}, or {\it MIPSGAL}. For this reason, we
suspect this source may be extra-galactic.

\paragraph{IRAS 18370-0607}
From the {\it GLIMPSE}-8\microns\ image, this object and the object
just to the north-east appear to be either post-merger galaxies, or
pinwheel nebulae as seen in interacting binary systems. That there are
two such objects close together seems to favour the former
explanation, although the extinction through this region of the Galaxy
would mean that these objects were extremely intrinsically bright. For
now we draw no definite conclusions as to the nature of these objects.

\paragraph{G026.10-00.07}
This source was observed in the radio survey of \citet{Downes80}, who
measured radial velocities of 33\kms\ from H110$\alpha$ and 104\kms\
from H$_{2}$CO. It would seem likely that the object as seen in Fig.\
\ref{fig:comp} is a foreground H{\sc ii}-region, and the H$_{2}$CO
beam was contaminated by emission from the RSGs. 

\citet{Wink82} determine that for near- and far-side kinematic
distances of the object, using the velocity measured by
\citet{Downes80}, the rate of Lyman continuum photons absorbed is
$\log (N_{\rm Ly}/s)$ = 48.54/50.08. As the far-side distance would
imply an extraordinarily-massive cluster of $\sim$8 O3 stars
\citep[e.g.][]{Sternberg03}, and as no obvious central cluster is
seen, the nearside distance seems more likely.

\paragraph{Star 1, Star 49, and IRAS 18364-0605}
It can be seen from Fig.\ \ref{fig:seds} that Star 49 has remarkable
near- and mid-IR excess. In addition, the IRAC images show the star
apparently at the centre-of-curvature of a bow-shock structure,
identified in the {\it IRAS} point-source catalogue as IRAS
18364-0605. Star~49 itself is one of the most luminous stars in the
cluster, while having an earlier-than-average spectral-type (K3). The
star's temperature places it close to the Yellow Hypergiants, a short
evolutionary phase experienced by stars on their way from the RSG to
the LBV/WR stages \citep{dJ98}. In clusters containing so many RSGs,
it is not unreasonable to expect to find one such object (see also
Star 15 in RSGC1, FMR06).

Star~49 is not as hot as the YHGs, however the large IR-excess is
suggestive of large amounts of warm circumstellar dust ejected in a
high mass-losing episode, possibly a precursor to blue-ward
evolution. It would be interesting to measure the mass-loss rate of
this object, and compare it to that of outburst of the YHG
IRC~+10~420, during which the mass-loss rate is inferred to have
reached 5$\times$10$^{-4}$\msunyr\ \citep{Oudmaijer96}. Since evolved
then the star has apparently evolved to an A-type supergiant
\citep{Klochkova02}, though this may be due to the dissipation of the
pseudo-photosphere created by the dense wind \citep{Smith04}.

Star~1 is by far the brightest object in this field in the $K$-band
($K_{S}$=2.9), and is highly reddened ($H-K_{S}$=1.798). It is not
possible to fit this star with a standard reddening law, assuming the
late spectral type of M5-6 derived from its CO-bandhead absorption. It
is likely this object has significant IR excess, possibly due to an
extreme mass-losing episode. Its radial velocity is $\sim$20\kms\
below that of the rest of the `cluster' stars, hence it is unlikely to
be a foreground giant. Indeed, the star may be part of the RSGC2
cluster, and its observed radial velocity offset by an expanding
optically-thick envelope; the velocity difference of $\sim$20\kms\ is
a typical outflow-speed for a RSG. It is possible that this star is an
extreme Red Hypergiant, such VY CMa -- a star with large IR excess and
inferred mass-loss rate of $\sim$2$\times$10$^{-4}$\msun\
\citep{Danchi94}. 

The Red- and Yellow-Hypergiants are extremely rare objects, however it
is not clear whether this is due to the exceptional nature of certain
stars, or whether all stars of a particular initial mass-range and
metallicity will pass through brief but extreme mass-losing episodes
such as these. If the latter is true, it is then not unreasonable to
expect to find such stars among the two Scutum-Crux RSG clusters, as
RSGC1 \& 2 appear to be of just the right age and initial
mass. Stars~1 and 49 certainly warrant further study in the context of
the evolution of RSGs at Galactic metallicity, as they appear to be on
the verge of shedding their outer layers and evolving blueward toward
the LBV/WN phases.

\subsection{The RSG clusters in the broader context of astrophysics}

From the cluster mass and age derived here, RSGC2 joins the nearby
RSGC1 (FMR06), Westerlund~1 \citep[Wd~1, ][]{Clark05}, the Arches
\citep{Figer02}, Quintuplet \citep{Figer99}, and Galactic centre
\citep[GC, ][]{Figer04} clusters, in a growing list of Galactic
analogues to Super Star Clusters. These objects represent ideal
natural laboratories in which to study the evolution of massive
stars. The Arches cluster is massive enough and young enough to
contain main-sequence O-stars up to the mass of $\sim$150\msun
\citep{Figer02}, whilst the age and mass of Wd~1 is tuned in such a
way that it contains 24 WRs -- 8\% of all those known in the Galaxy
\citep{C-N02,N-C05,Groh06,Crowther-Wd106}. Meanwhile \citet{Martins07}
recently applied abundance analysis to the unusually-large number of
Opfe/WN9 stars in the GC cluster to tie down their evolutionary
status.

Of this collection of massive Galactic clusters, the two RSG clusters
are evidently the elder, not only from the population-synthesis
analysis presented here and in FMR06, but also from the lack of
diffuse radio emission at the centre of the clusters, and hence of few
remaining hot main-sequence stars (we note that, while RSGC2 has a
radio nebula just to the south-west, radio recombination line
observations place this object in the foreground -- see Sect.\
\ref{sec:others}).

This gives the two clusters a unique role in the context of massive
stellar evolution, as they offer the opportunity to study a
statistically-significant population of RSGs and probe the evolution
of stars in the mass-range of $\sim$15-25\msun. Evidence is growing
that such stars are the progenitors of Type-II SNe \citep{vD03,
Smartt04}, while they may contribute significantly toward
Galactic-scale dust production, particularly in low-metallcity
starbursts where Carbon-sequence WRs are absent and AGB stars are yet
to form \citep[see discussion by][]{Massey05}.

As hinted by the distribution of spectral-types (Sect.\
\ref{sec:spectypes}), the clusters likely have similar abundances,
representative of the rest of the Galaxy. In addition, the fact that
there are {\it two} clusters, with slightly different ages -- and
hence initial masses of RSGs -- now permits evolutionary studies at
uniform metallicity, and as a function of initial mass.

\subsubsection{The RSGCs as a probe of the Galactic Z-gradient}
Aside from the unusually-large number of RSGs and the opportunities
they present for studying stellar evolution, an additional interesting
aspect of the clusters is their location in the Galactic plane at a
Galacto-centric distance of $\sim$4kpc. Here, they are close to where
the disk meets the central bulge, within the proposed `ring' of
enhanced stellar density \citep{Bertelli95}. 

Chemical abundance analyses of this region could be key to
understanding the formation and evolution of our Galaxy, in the
transition zone between the Galactic disk and bulge. An important
constraint on models of the formation of the Galaxy and its central
bulge is the radial metallicity gradient. Abundance analyses of H{\sc
ii} regions, planetary nebulae, and early-type stars have found a
steadily increasing metal content from 18 -- 5kpc
\citep[e.g. ][]{Afflerbach97,M-Q99,Rolleston00}. The metallicity
within 5kpc of the Galactic centre is less clear: \citet{Smartt01}
find that the metallicity continues to increase at the same slope down
to 2.5kpc, although curously not for oxygen; while studies of the
inner $\sim$50pc have have revealed roughly solar abundances
\citep[e.g. ][]{Ramirez00,Najarro04}.

The RSGCs are now a powerful tool with which to probe what could
potentially be the transition zone between the Galactic disk and
bulge, and the location where the metallicity gradient breaks
down. Near-IR spectral analyses of RSGs can yield Fe abundances
\citep[e.g.\ ][]{R-O05}; while analysis of the LBV candidate, and any
BSGs in the clusters, would provide direct measurements of Fe and
$\alpha$-group elements such as Si and Mg \citep{Najarro06}.


\section{Conclusions}
Using near-IR spectroscopy, and 2MASS/GLIMPSE/MSX photometry, we have
shown that there is a second reddened, massive young cluster of RSGs
in the Galactic plane at $l=25-26$\degr. We find that this cluster,
RSGC2, contains 26 RSGs, almost twice as many as the nearby
RSGC1. From evolutionary synthesis and kinematic measurements we infer
that RSGC2 is slightly older and the more massive of the two, with a
mass comparable to that of Westerlund~1. Together, the two Scutum-Crux
RSG clusters harbour $\sim$20\% of all known RSGs in the Galaxy, and
now offer an unprecedented opportunity to study pre-supernova
evolution at uniform metallicity. Further, new infrared/radio survey
images reveal several background RSGs, and candidates for a supernova
remnant and a Luminous Blue Variable in the field of RSGC2. Along with
the proximity of RSGC1, this suggests intense, recent, region-wide
star-formation activity at the point where the Scutum-Crux Galactic
arm meets the Galactic bulge. Future abundance studies of this region
would yield important information in the study of the Galactic
metallicity gradient, and the interaction between the disk and the bulge.

\acknowledgments We would like to thank the anonymous referee for a
careful reading of the manuscript and several suggestions which
improved the paper. For useful discussions we thank Simon Clark, and
Tom Jarrett for discussions concerning the nature of IRAS
18370-0607. The material in this work is supported by NASA under award
NNG~05-GC37G, through the Long-Term Space Astrophysics program. IRMOS
is supported by NASA {\it James Webb Space Telescope}, NASA Goddard
Space Flight Center, STScI DDRF, and KPNO. This research has made use
of the {\sc simbad} database, Aladin \& IDL software packages, and the
GSFC IDL library.

\bibliographystyle{/fat/Data/model_paper/aa}
\bibliography{/fat/Data/model_paper/biblio}

\appendix
\section{Appendix}

\begin{figure}[h]
  \centering
  \includegraphics[width=15cm,bb=15 15 556 570]{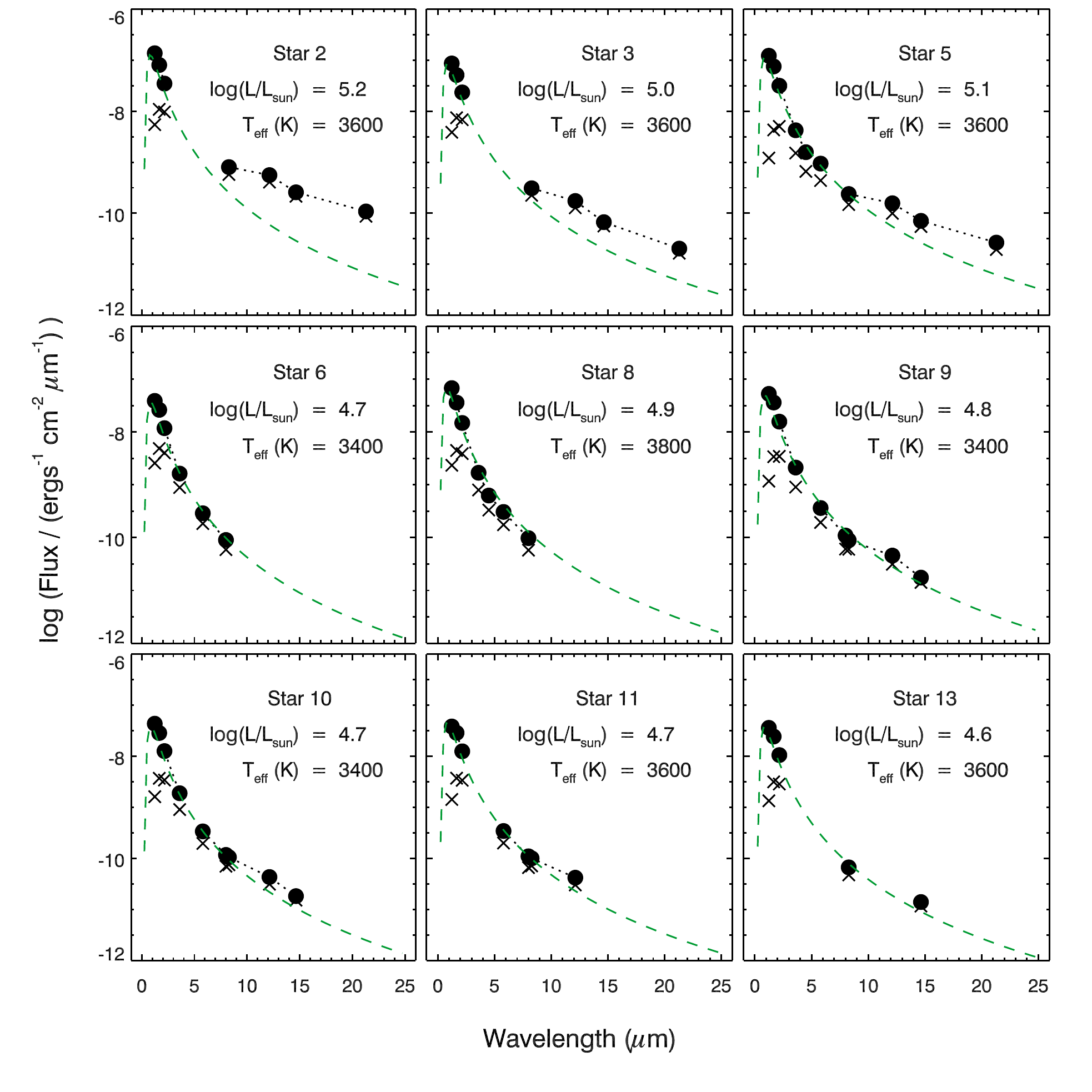}
  \caption{Spectral energy distributions of the cluster members, using
  data from the point-source catalogues of {\it 2MASS}, {\it GLIMPSE}
  and {\it MSX}. The raw photometry is plotted as crosses, and the
  de-reddened photometry as filled circles. Neither {\it GLIMPSE} data
  below 5$\sigma$, nor {\it MSX} upper-limit data are plotted. The
  green dotted-line in each panel represents a black-body curve
  appropriate for each star's $K_{S}$-band magnitude, reddening and
  temperature, and the kinematic distance to the cluster. We note that
  there may be some contamination in the {\it MSX} aperture between
  Stars 2 and 6, which may explain the large mid-IR excess of Star 2.}
  \label{fig:seds}
\end{figure}
\addtocounter{figure}{-1}
\begin{figure}[p]
  \centering
  \includegraphics[width=15cm,bb=15 15 556 570]{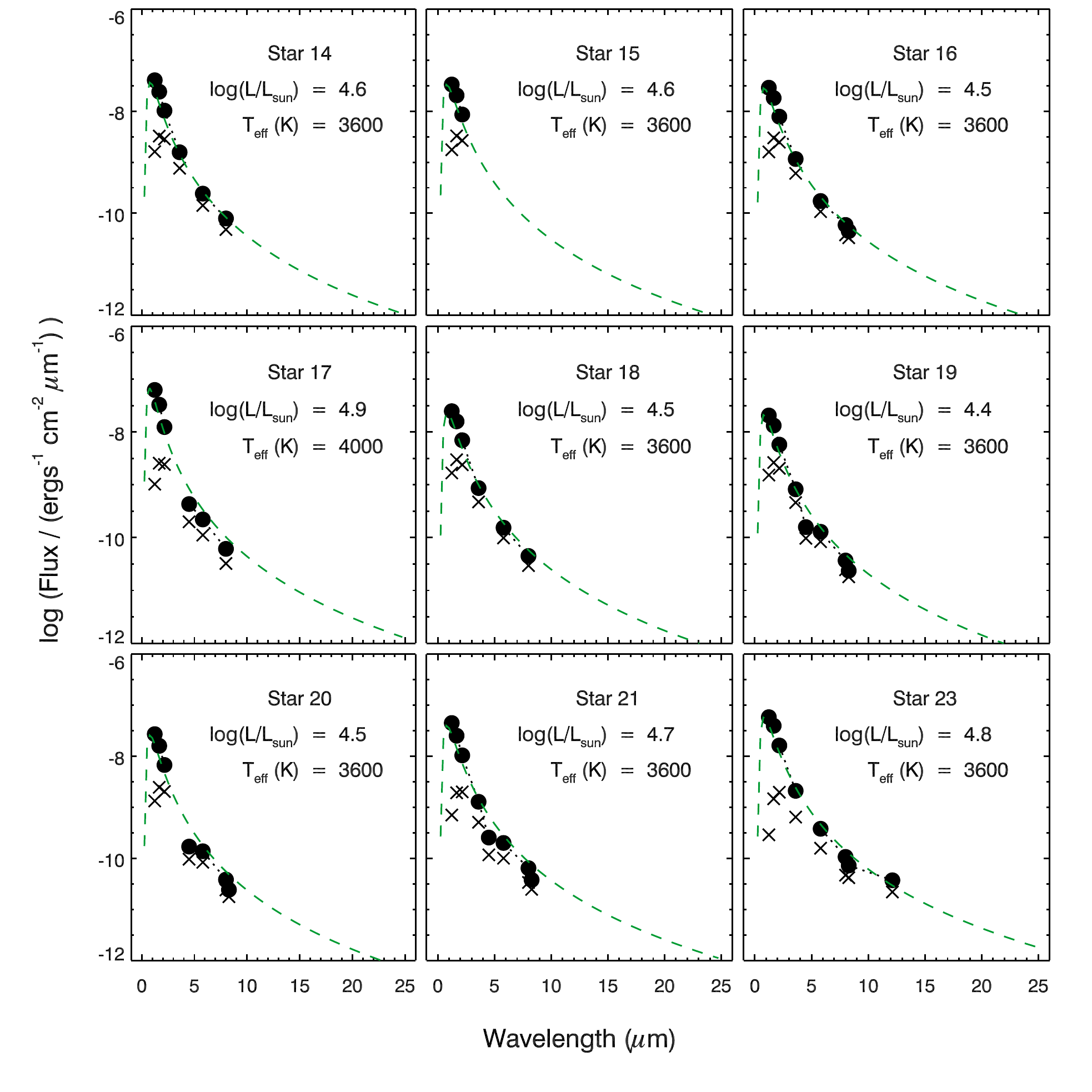}
  \caption{cont.}
\end{figure}
\addtocounter{figure}{-1}
\begin{figure}[p]
  \centering
  \includegraphics[width=15cm,bb=15 15 556 570]{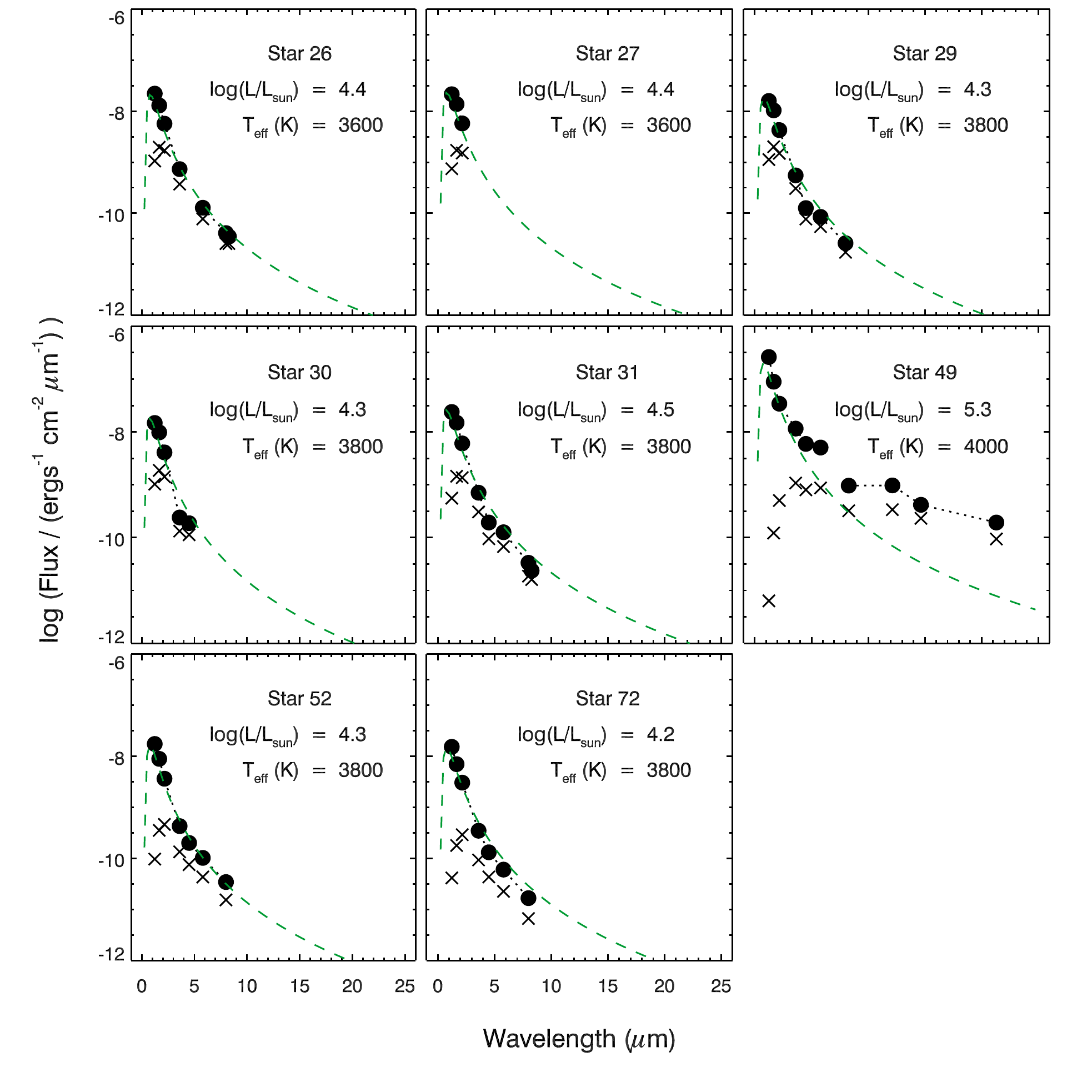}
  \caption{cont.}
\end{figure}

\end{document}